\documentclass[aps,preprint,amssymb,12pt,floatfix]{revtex4}
\pdfoutput=1
\usepackage{lscape,graphicx}
\usepackage{rotating}
\usepackage{epstopdf}
\usepackage{color}
\usepackage{subfig}
\usepackage{wrapfig}

\begin{document}
\title{Capturing the essence of folding and functions of biomolecules using Coarse-Grained Models}
\author{Changbong Hyeon}
\affiliation{School of Computational Sciences, Korea Institute for Advanced Study, Seoul 130-722, Republic of Korea}
\author{D. Thirumalai}
\thanks{Email: thirum@umd.edu}
\affiliation{Biophysics Program, Institute For Physical Science and Technology, University of Maryland, College Park, MD 20742, USA}
\date{\today}
\begin{abstract}
The distances over which biological molecules and their complexes can function range from a few nanometres, in the case of folded structures, to millimetres, for example during chromosome organization. Describing phenomena that cover such diverse length, and also time scales, requires models that capture the underlying physics for the particular length scale of interest. Theoretical ideas, in particular, concepts from polymer physics, have guided the development of coarse-grained models to study folding of DNA, RNA, and proteins. More recently, such models and their variants have been applied to the functions of biological nanomachines. Simulations using coarse-grained models are now poised to address a wide range of problems in biology. 
\end{abstract}
\maketitle

Minimal models that capture the essence of complex phenomena has a rich history in the natural sciences. 
In condensed matter physics insights into many phenomena have emerged from analytic theories of models, which use effective many body Hamiltonians that succinctly capture  the essence of the problems \cite{Anderson:1997}.  
Examples include phase transitions, superfluidity and superconductivity. 
However, complex problems, such as spin glasses \cite{Mezardbook},  structural glasses \cite{Kirk89JPhysA,Vas07ARPC} and a host of problems in biology such as protein and RNA folding and functions of macromolecules have resisted solutions using purely theoretical methods. 
These and other  problems in material science in which a wide range of time, energy, and length scales are intertwined require well-designed computer simulations, which capture the essential  features of the systems.   
Although the temptation to use detailed atomic simulations in protein folding and more complicated problems is hard to resist, such an approach has given us only limited insights.  
In contrast, since the first classical molecular dynamics simulation that reported phase transition in hard-sphere systems \cite{Alder57JCP}, it has been clear that coarse-grained (CG) models are often the only way to describe phenomena that involve an interplay of multiple energy and time scales. 
Nowhere is the need for CG models greater than in biology in which self-assembly of macromolecules and their functions, which involve multiple partners, occur on time and length scales that cover many orders of magnitude. 
In the context of protein and RNA folding,  simulations using CG models, guided by theoretical concepts \cite{Hyeon05BC,OnuchicCOSB04,Shakhnovich06Chemrev,Dill08ARB,Thirumalai10ARB}, have unearthed the principles of self-assembly.  
More recently models, which were introduced to describe folding of isolated proteins \cite{Levitt75Nature,Honeycutt90PNAS,Clementi00JMB,Karanicolas2002ProtSci,Weinkam10ACR} and RNA \cite{Hyeon05PNAS}, have also been adopted and extended in novel ways to predict functions of large complexes such as  ribosomes \cite{Whitford10RNA}, molecular chaperones \cite{Hyeon06PNAS}, enzyme catalysis \cite{Pisliakov09PNAS}, protein insertion into membranes \cite{Rychkova2010PNAS} and a number of motors \cite{Strajbl2003PNAS,Koga06PNAS,Hyeon07PNAS,Hyeon07PNAS2,Takano10PNAS,Tehver10Structure,Liu09PNAS,Kamerlin11ARPC}. 
These developments have resulted in a quiet revolution, which has provided molecular insights into a variety of biological processes.

In the last two decades, fundamental breakthroughs into structural organization and dynamics of proteins, RNA, and DNA have been achieved using theoretical concepts from polymer physics \cite{deGennesbook} and CG simulations.
Here, we describe how simulations using a variety of CG models  have been successful in describing dynamical processes in biology spanning a wide range of length scales. 
These achievements have been further extended to probe folding under cellular conditions \cite{Minton08ARB,Elcock10COSB,Cheung05PNAS}, and more recently to  describe functional dynamics of biological nanomachines \cite{Koga06PNAS,Hyeon06PNAS,Hyeon07PNAS,Hyeon07PNAS2,Tehver10Structure}. 
The use of CG models and simple theoretical ideas have also found fruitful applications in many other areas in biology such as gene networks, systems biology, and analysis of complex metabolic pathways.

\section*{Length scales determine extent of coarse-graining.}
Description of reality using models requires a level of abstraction, which depends on the phenomenon of interest. For example, near a critical point, exponents that describe the vanishing of order parameter or divergence of correlation length are universal, depending only on the dimensionality ($d$), and are impervious to  atomic details.  
These findings, which are rooted in the concepts of universality and renormalization group \cite{Fisher74RMP},  are also applicable to the properties of polymers  \cite{deGennesbook}.
For example, the size of a long homopolymer and its distribution of end-to-end distance depend only on the solvent quality, the degree of polymerization, and $d$, but not on the details of monomer structure \cite{deGennesbook}. However, on length scales that are on the order of a few nm one has to contend with chemical properties of the monomer.  

In the absence of rigorous theoretical underpinnings, intuitive arguments and phenomenology come into play in modeling complex biological processes (Box 1). Here, also the level of description depends on length scales. 
In nucleic acids, at a short length scales ($l\lesssim 5$ \AA) detailed chemical environment determines the basic forces (hydrogen bonds and dispersion forces) between two nucleotides. 
On the scale $l\sim (1-3)$ nm interactions between two bases, base stacks and grooves of the nucleic acids become relevant. Understanding how RNA folds ($l\sim (1-3)$ nm) requires energy functions that provide at least a CG description of nucleotides, and interactions between them in the native state and excitations around the folded structure. 
On the persistence length scale $l_p\approx 150$ bp $\approx 50$ nm \cite{Bustamante94SCI} and beyond it suffices to treat dsDNA as a stiff elastic filament without explicitly capturing the base-pairs.  
If $l\sim \mathcal{O}(1) $ $\mu m$ dsDNA behaves like a self-avoiding polymer \cite{VallePRL05}. 
On the scale of chromosomes ($l\sim mm$) a much coarser description suffices. 
Thus, models for DNA, RNA, and proteins vary because the scale of structural organization changes from nearly mm in chromatin to several nm in the folded states of RNA and proteins.

\section*{Polymer models for dsDNA and chromosome structure.}

Length, $L$, of double stranded DNA (dsDNA) exceeds a few $\mu m$ with persistence length, $l_p \approx$ 50nm.  On these scales global properties of dsDNA, such as the end-to-end distance and the dependence of $l_p$ on salt concentration, are not greatly affected by fluctuations of individual base pairs. Consequently, dsDNA can be treated as a fluctuating elastic material, for which the Worm-like Chain (WLC) is a suitable polymer model. On  much longer scales ($L \sim$ 1mm), which is relevant to chromosome, the genomic material can be described as a flexible polymer. Using these scale-dependent models a number of predictions for DNA organization and dynamics can be made.  \\

{\bf Looping dynamics:} 
Loop formation in biopolymers is an elementary process in the self-assembly of DNA, RNA and proteins. 
However, understanding cyclization kinetics is complicated because multiple length scales and internal chain modes are intertwined in bringing distant parts of DNA into proximity. 
For a short chain, the cyclization time, $\tau_c$ scales as $L^{3/2}$ while $\tau_c\sim L^2$ when $L$ increases \cite{PastorJCP96,Toan08JPCB}. 
The problem of cyclization becomes more challenging in the looping dynamics of dsDNA, an elementary process that is relevant in controlling gene expression and DNA condensation.  
In the CG model a single-pitch of a double helix, formed by 10.5 base-pairs, represents one interaction center (Fig.\ref{DNA}a). 
Thus, $l_p$ encompasses $(14-15)$ CG interaction centers ($l_p\approx$ 150 bp).
The parameters for bond and bending potentials along the chain, consisting of multiple CG centers, are selected to reproduce the persistence length of dsDNA \cite{VologodskiiMacro00,Vologodskii92JMB}, allowing us to study various dynamics of dsDNA, stretching, looping, or supercoiling from the perspective of polymer physics.   
The ease of  loop formation and the associated kinetics is characterized by  $L/l_p$. 
For $L/l_p\sim \mathcal{O}(1)$, energy required to bend dsDNA makes the cyclization  difficult for short chains. 
In contrast, when $L/l_p\gg 1$ the cyclization between two ends gets harder because of loss of chain entropy.  
Theory and simulations using CG model showed that $\tau_c$ is the shortest when $L/l_p\approx 2-3$ \cite{VologodskiiMacro97,HaELett03,HyeonJCP06} (Fig.\ref{DNA}a). 
Interestingly, in looping of dsDNA responsible for gene regulation in prokaryotes $L\approx 100$ bp ($L/l_p\approx 0.7$). 
For such a dsDNA with $L\approx 100$ bp sequence effects are also relevant \cite{WidomMC04,VologodskiiPNAS05,Savelyev10PNAS}. \\

{\bf Stretching dsDNA:} 
Fluctuations of dsDNA on scales comparable to $l_p$ can be described using WLC model, which parameterizes dsDNA as a polymer which resists bending on scale $\sim l_p$.  Smith \emph{et. al.} measured the response of a 97 kbp dsDNA (countour length $L\approx 33.0$ $\mu m$) from $\lambda$-phage to a stretching force, $f$ \cite{BustamanteSCI92,Bustamante94SCI} (Fig.\ref{DNA}b). 
In the absence of $f$, $\lambda$-DNA conformations are determined by thermal fluctuations, whereas loss in chain entropy must be overcome to stretch  dsDNA  $f \ne 0$. 
The free energy of stretching of a semiflexible chain under tension is equivalent to a quantum mechanical problem of a dipolar rotor with moment of inertia $l_p$ in an electric field $f$.  An extrapolation formula obtained by numerically solving the quantum mechanical problem that accurately describes the measured force as a function of extension (Fig.\ref{DNA}b).  Fits to experimental data yield  $L$ of $\lambda$-DNA ($32.80\pm 0.10$) $\mu m$ and $l_p\approx (53.4\pm 2.3)$ nm, thus confirming most directly that dsDNA is a semiflexible chain.  \\

\textbf{Confined polymers and bacterial chromosome segregation:} Replication and passage of genetic information to daughter cells are major events in cell reproduction. 
These complex events are remarkably accurate even in simple organisms. 
Although chromosome segregation is likely to be complex and well orchestrated, it has recently been proposed that confinement-induced entropic forces due to  restrictions in cellular space is sufficient to drive chromosome segregation in bacteria \cite{jun06PNAS,jun2010NRM} (Fig.\ref{DNA}c). 
This proposal was  formulated using  molecular simulations of tightly confined self-avoiding polymers chains in cylindrical space, which show that the chains segregate and become spatially organized reminiscent of that observed in bacteria. In such highly confined spaces  polymer conformations are determined by $\xi$, the size of a renormalized  structural unit, the Flory radius $R_F$ in the absence of confinement, and the length ($P$) and diameter $D$ of the cylinder. In \textit{E. Coli.} the values are $\xi$ = 87 nm, $R_F$ = 3.3 $\mu m$, $D$ and $P$ are 0.24 $\mu m$ and 1.3 $\mu m$, respectively. 
Armed with the results for confined polymers, a concentric shell model for bacterial chromosome was proposed \cite{jun06PNAS,jun2010NRM} in which the nucleoid was modeled as an inner and outer cylinder. 
The unreplicated ``mother" strand, a self-avoiding chain, is restricted to the inner compartment whereas the ``daughter" chain (obtained in simulations by adding monomers at a set time in the Monte Carlo simulations) are free to explore the entire nucleoid volume. The results of the simulations show that the newly added (or replicated) chain segregates to the periphery of the nucleoid, driven by gain in entropy, and become spatially organized as they are synthesized (Fig.\ref{DNA}c).  CG modeling combined with polymer theory lead to the discovery that entropic forces alone are sufficient to drive chromosome segregation in bacteria, with proteins perhaps playing a secondary role in poising the state of the chromosome for enabling the entropy-driven mechanism.   \\

\textbf{Chromosome Folding:} In eukaryotic cells chromosomes fold into globules that spatially occupy well-defined volumes known as chromosome territories \cite{cremer2001NRG}.  
In this process widely separated gene-rich regions are brought into close proximity. Knowledge of the spatial arrangement of chromosomes is important in describing gene activity and the state of the cell. Polymer physics concepts have  been used to describe the structures of folded chromosome using constraints derived from experiments. These calculations  have provided considerable insights into their compartmentalization in the nucleus \cite{grosberg1993EPL}.  
A number of models, such as the random walk model, and models that connect mega-based size domains by chromatin loops have been used to describe higher structures of chromatin. 
The experimental resolution is roughly 1Mb ($\approx$ 340$\mu m$), and consequently coarse-graining in this context must be on  length scales on the order of a $\mu m$.   
Recently folding principles for human genome were proposed using data for long-range contacts between distinct loci as constraints \cite{lieberman09Science}. 
Experiments showed that contact probability, $I(s)$, between loci in a chromosome, which is separated by genomic distance $s$ (measured in units of bp) exhibits a power law decay in the range $\sim$ 500 kb to $\sim$ 7 Mb. The observed dependence  $I(s) \sim s^{-1}$ can be rationalized using polymer models (Fig.\ref{DNA}d) introduced a number of years ago in describing collapse of homopolymers \cite{grosberg1988JP}. 
If chromosome folds up into an equilibrium globule (polymer in a poor solvent) then $I(s) \sim s^{-1.5}$, which cannot account for the experimental observations. An alternate model suggests that interface DNA can organize itself  into a fractal globule, which is compact and not entangled as an equilibrium globule would be.  
Monte Carlo simulations of a polymer with 4000 beads (1 bead = 1200 bps $\sim 0.4 \mu m$) were used to generate conformations of fractal and equilibrium globules. 
The power law decay of $I(s)$, with exponent $\approx -1$, is consistent with measurements (Fig.\ref{DNA}d). 
More importantly, the unknotted fractal globules loci that are close in genomic sequence are also in proximity in three dimensional spatial arrangement, which clearly is relevant for gene activity.  

\section*{RNA Folding}

Since the discovery that RNA can serve as enzymes there has been great impetus to describe their folding in quantitative terms. RNA folding landscape is rugged because of interplay of several competing factors. 
First, phosphate groups are negatively charged, which implies that polyelectrolyte effects  oppose folding. 
Valence, size and shape of counterions, necessary to induce compaction and folding \cite{ThirumARPC01}, can dramatically alter the thermodynamics and kinetics of RNA folding. 
Second, the nucleotides purine and pyrimidine bases have different sizes but are chemically similar. 
Third, only $\sim 46$ \% of bases form canonical Watson-Crick base pairs while the 
remaining nucleotides are in non-pairing regions \cite{DimaJMB05}. 
Fourth, the lack of chemical diversity in the bases results in RNA easily adopting alternate misfolded conformations, which means that the stability gap between the folded and misfolded structures is not too large. 
Thus, the homopolymer nature of the RNA monomers, the critical role of counterions in shaping the folding landscape, and the presence of low-energy excitations around the folded state make RNA folding a challenging problem \cite{Hyeon05BC}. \\

{\bf Polyelectrolyte (PE) effects:}
To fold, RNA must overcome the large electrostatic repulsion between the negatively charged phosphate groups. 
PE based theory shows that multivalent cations ($Z>1$) are more efficient in neutralizing the backbone charges than monovalent ions - a prediction that is borne out in experiments. The midpoint of the folding transition $C_m$, the ion-concentration at which the folded and unfolded states are equal, for \emph{Tetrahymena} ribozyme is $\sim 3\times 10^6$ fold greater in Na$^+$ than in cobalt-hexamine ($Z=3$)! 
The nature of compact structures depends on $Z$ with the radius of gyration scaling as  $R_G\propto 1/Z^2$, which implies compact intermediates have larger free energy as $Z$ increases.  
Thus, folding rates should decrease as $Z$ increases, which also accords well with experiments \cite{moghaddam09JMB}. 
Polyelectrolyte theory also shows that counterion charge density $\zeta=Ze/V$ should control RNA stability. 
As $\zeta$ increases, RNA stability should increase - a prediction that was validated using a combination of PE-based simulations and experiments. 
The changes in stability of in \emph{Tetrahymena} ribozyme in various Group II metal ions (Mg$^{2+}$, Ca$^{2+}$, Ba$^{2+}$, and Sr$^{2+}$) showed a remarkable linear variation with $\zeta$ \cite{Koculi07JACS}. 
The extent of stability is largest for ions with largest $\zeta$ (smallest $V$). 
Brownian dynamics simulations showed that this effect could be captured solely by non-specific ion-RNA interactions \cite{Koculi07JACS}. 
These findings and similar variations of stability in different sized diamines show that (i) the bulk of the stability arises from non-specific association of ions with RNA, and (ii) stability can be greatly altered by valence, shape, and size of the counterions. 
 \\

{\bf Structures of RNA intermediates:} 
The complete characterization of counterion-mediated RNA folding requires structural description of the unfolded ($U$), intermediate ($I$), and the folded states. 
Structures of the folded states can be obtained using crystallography or NMR. However,  it is difficult to characterize the ensemble of structures populated at low ($U$) and moderate ($I$) ion concentrations (C). 
To obtain the ensemble of $I$ structures from time-resolved SAXS data, 
a CG model for \emph{Tetrahymena} group I was constructed by representing (5-6) nucleotide pairs by a single sphere (Fig.\ref{RNA}a) \cite{RussellPNAS02}.  
The salient findings are: 
(i) At times prior to global collapse the domains of the ribozymes are extended because PE effects dominate. 
(ii) On time scales that are much less than the overall folding time there is a drastic reduction in the size of RNA. 
The folding intermediates are fluid-like and must be a mixture of species that contain specifically collapsed structures (large degree of native-like order) and non-specifically collapsed conformations (low degree of native-like order). \\

{\bf Complexity of hairpin formation: }
When viewed on length scales that span several bps folding of a small RNA (or DNA) hairpin is remarkably simple. 
However, when probed on short times ($ns-\mu s$ range) the formation of a small hairpin involving turn formation and base-stacking is remarkably complex. 
Recent experiments show that the kinetics of hairpin formation in RNA (or ssDNA) deviates from the classical two-state kinetics and is best described as a multi-step process \cite{Ma06JACS}. 
Additional facets of hairpin formation have been revealed in single molecule experiments that use mechanical force ($f$). These experiments prompted simulations that vary both $T$ and $f$.
The equilibrium phase diagram showed two basins of attraction (folded and unfolded) at the locus of critical points $(T_m, f_m)$. At $T_m$ and $f_m$ the probability of being unfolded and folded is the same.
The free energy surface obtained from simulations  explained the sharp bimodal transition between the folded and unfolded state when the RNA hairpin is subject to $f$ \cite{Hyeon05PNAS,Hyeon08JACS}. 
Thus, from thermodynamic considerations, hairpin formation can be described as a two-state system. 

Upon temperature quench hairpin forms by multiple steps \cite{Hyeon08JACS}  as observed in the recent kinetic experiments. 
Folding pathways between $T$-quench and $f$-quench refolding are different(see Fig.\ref{RNA}b). 
The initial conformations generated by forced-unfolding are fully extended. They are structurally homogeneous.
The first event in folding upon $f$-quench  is loop formation, which is a slow slow nucleation process(see Fig.\ref{RNA}b). 
Zipping of the remaining base pairs leads to rapid hairpin formation.
Refolding upon $T$-quench commences from a structurally broad ensemble of unfolded conformations. 
Therefore, nucleation can originate from many regions in the molecule(see Fig.\ref{RNA}b). 
The simulations showed that the complexity of the folding landscape observed in ribozyme experiments was already reflected in the formation of simple RNA hairpin \cite{Chen00PNAS,Hyeon05PNAS} just as $\beta$-hairpin formation captures much of the complexity of protein folding \cite{ThirumCOSB99}.

 \section*{Protein Folding}

The impetus to understand the mechanisms of protein folding comes from a number of different sources. First, there is increasing need to produce models that can predict folding thermodynamics and kinetics at conditions used in experiments. Second, it is urgent to describe the biophysical basis of misfolding and the link to neurodegenerative diseases. Third, as we move towards a system level description of cellular processes it is important to develop theoretical models for describing folding in crowded solutions as well as folding of proteins as they are synthesized by the ribosome. \\

{\bf Molecular Transfer Mode (MTM):} The validity of models can only be assessed by  comparing simulation results (obtained under conditions used in experiments)  to experiments.  Majority of computational studies use temperature to trigger folding and unfolding whereas a substantial number of experiments use denaturants for the same purpose, thus making it difficult to validate  the models.   This difficulty has been overcome with the introduction of a phenomenological  MTM \cite{Obrien08PNAS,liu2011PNAS}, which combines simulations performed in condition A (for example fixed temperature, $T_1$ and zero denaturant concentration) and the sampled conformations are assigned appropriate Boltzmann weight  such that the behavior in solution condition B  ($T$ and non-zero denaturant concentration for example) can be accurately predicted without running additional simulations. The MTM theory shows that this procedure is exact provided the conformations of the protein are exhaustively sampled in condition A, and is only limited by the accuracies of the force fields.  Applications of MTM requires the free energy cost of transferring a given protein conformation from A$\rightarrow$B, which were taken from experimentally measured transfer free energies for the peptide backbone and each amino acid.  The MTM simulations for protein L (an $\alpha/\beta$ protein) and the nearly all $\beta$-sheet cold shock protein quantitatively reproduced measured values of the dependence of the population of the folded state as a function of denaturants (Fig.\ref{ProteinsFig}a). Surprisingly, MTM-based simulations also accurately predicted denaturant-dependent measurements in single molecule experiments.  \\

{\bf Mechanical force in protein folding:} A number of single molecule experiments, which use mechanical force ($f$)  in various modes (force ramp, force quench, and constant force) to initiate folding from arbitrary regions in the energy landscape, have given a new perspective on protein (and RNA) folding \cite{tinoco2006QRB,Fernandez04Science}.  These experiments, which monitor time-dependent changes in the extension, $x(t)$, of the protein of interest showed that folding occurs in multiple stages upon force quench. 
The power of  these experiments are fully realized only by combining them with theory \cite{Hyeon08PNAS,HyeonMorrison09PNAS,Dudko06PRL,Hyeon03PNAS} and simulations.  Such an approach was used to construct the folding landscape of the nearly 250-residue green fluorescent protein (GFP), which has a barrel-shaped structure consisting of 11 $\beta$-strands with one $\alpha$-helix in the C-terminus. Using simulations with self-organized polymer (SOP) representation \cite{Hyeon06Structure} of GFP at the loading rate used in experiments a rich and complex folding landscape was predicted (Fig.\ref{ProteinsFig}b).  
Unfolding of the native ($N$) began with rupture of the $\alpha$-helix leading to [GFP$\Delta \alpha$] intermediate. Subsequently, there was a bifurcation in the unfolding pathways.  In most cases, the route to the unfolded ($U$) involved population of two additional intermediates, [GFP$\Delta\alpha\Delta \beta_1$] ($\Delta\beta_1$ represents forced-rupture of N-terminal $\beta$-strand) and [GFP$\Delta\alpha\Delta\beta_1\Delta\beta_2\beta_3$]. The most striking prediction of the simulations was that  the minor pathway had only one  intermediate [GFP$\Delta\alpha\Delta\beta_{11}$] besides [GFP$\Delta\alpha$]. The predictions using SOP simulations of GFP were quantitatively validated by single molecule experiments \cite{Mickler07PNAS}. \\

{\bf Cotranslational folding:} With the determination of the ribosome structures \cite{YonathCell01,SteitzARB03} there is great interest in the folding of proteins as they are synthesized. 
Upon synthesis, which occurs at the rate of about 20 amino acids per second in \textit{E. Coli.}, the polypeptide chain traverses a roughly cylindrical tunnel whose lining changes from the peptidyl transfer center (PTC) to the exit that is $\sim$ 10 nm from PTC (Fig.\ref{ProteinsFig}c). 
Experiments have shown that it is likely that certain regions can accommodate $\alpha$-helices depending on the sequence, which is of particular interest for transmembrane helices that can be directly inserted into the membrane by the translocon. 
Inspired by these experiments, theory and simulations were used to show that the extent of helix formation does depend on the sequence \cite{Ziv2005PNAS}, the diameter of the tunnel, and potential interactions between the nucleotides and residues that line the tunnel and the polypeptide chain.  

More recently, several experiments  have probed the possibility of tertiary structure formation especially in the vestibule near the exit tunnel, whose volume is large enough for tertiary structure formation of the N-terminal region of the protein. 
CG simulations, which use either $C_{\alpha}$ model \cite{Elcock2006PLOSCompBio} or $C_{\alpha}$-SCM \cite{obrien10JACS} and all atom representation of RNA or TIS or four site model for RNA, have been used to interrogate coupled-synthesis and folding (Fig.\ref{ProteinsFig}c).  
Some general results were found in these simulations. 
(1) Polypeptide synthesis and folding are not coupled for single domain proteins, which require the synthesis of complete protein for folding to commence. 
(2) However, cotranslational folding is prevalent in multi-domain proteins in which the N-terminus region is likely to fold as it exits the tunnel. In this case the \textit{in vivo} folding pathway is expected to be different than \textit{in vitro}. 
(3) Simulations also suggest that interaction with the ribosome surface decreases folding pathway diversity and results in a more compact transition state structure \cite{obrien10JACS}. \\

\section*{Towards folding under cellular conditions} 
Cellular interior is replete with a host of macromolecules, which  can alter all processes ranging from transcription to folding. For example, in \emph{E. coli} ribosome ($\sim 20.8$ nm), polymerases and other protein complexes occupy merely 22 \% of total volume and small complexes and other small complexes constitute about 8 \% of the total volume. 
Thus, unlike \emph{in vitro} experiments where folding is studied   in an aqueous solution corresponding to infinite dilution conditions, crowding effects have to be taken when describing their behavior \emph{in vivo}.  
A simple calculation shows that average spacing between cytoplasmic proteins is $\sim 4$ nm \cite{phillips2009physical} (or $\rho\gtrsim (50-400)$ mg/ml \cite{minton2001JBC}). 
Given that the diameter of a typical proteins ($\sim$ 300 aa) is $\sim 4$ nm, the cell is an extremely crowded place, which severely inhibits conformational fluctuations that are easily realized in typical \emph{in vitro} experiments.  

An approximate mimic of the cellular environment can be realized by adding high concentrations of natural or synthetic macromolecules. 
Consider the simplest case of a crowding agent with radius $R_c$ (Ficoll 70 for example) that is inert towards the protein or RNA. 
The volume fraction $\varphi_c=\rho v$, where $v(=\frac{4}{3}\pi R_c^3)$, the volume of the crowding particle, can be altered by $R_c$ even with $\rho$ fixed. 
The first crowding simulation used a $C_{\alpha}$-SCM of a $\beta$-sheet protein in the presence of spherical crowding particles with $0\leq \varphi_c\leq 0.25$ \cite{Cheung05PNAS}. 
The CG simulations showed that when only excluded volume interactions dominate stabilities of globular proteins relative to $\varphi_c=0$ \cite{Cheung05PNAS} are enhanced (Fig.\ref{crowding}a).  
The extent of stability change,  measured using $\Delta T_m=T_m(\varphi_c)-T_m(\varphi_c=0)$, showed that $\Delta T_m\sim \varphi_c^{1/3\nu}$ where the $\nu(=3/5)$ is the Flory exponent that characterizes the size of the unfolded states of proteins. 
The scaling of  $\Delta T_m(\varphi_c)$ with $\varphi_c$ has  been confirmed in recent experiments \cite{dhar10PNAS}. 
Simulations also showed that the folding rate $k_F(\varphi_c)$, is also affected when $\varphi_c\neq 0$. 
Rate, $k_F(\varphi_c)$, increases monotonically till an optimum value, and subsequently decreases (Fig.\ref{crowding}b). 
Interestingly, identical behavior was observed in the dependence of the relaxation rate of phosphoglycerate kinase (PGK) as a function of Ficoll concentration. 
The simulation results on crowding-induced effects on the smaller $\beta$-sheet WW domain explains several aspects of folding of PGK (Fig.\ref{crowding}c).

In an insightful application of simulations it was recently shown that crowding can alter catalytic activity of kinase (Fig.\ref{crowding}c) \cite{dhar10PNAS}. 
As is common in many kinases, PGKs that transfers phosphate group from diphosphoglycerate to ADP, has a catalytic site between the N- and C- lobes connected by flexible hinge.
To perform kinase activity, PGK must undergo a large scale structural movement that reduces the distance between N- and C-lobes.  
It was found that PGK activity is increased over 15-fold in 200 mg/mol ($\varphi_c\approx 0.2$)  Ficoll 70.
The enhancement in activity was attributed to crowding-induced shape change that brings the N and C lobes in proximity.

\section*{Biological nanomachines}

Biological machines are typically multi-subunit constructs that carry out myriads of functions by interacting with a range of proteins and RNA. 
Examples of such machines include molecular motors (kinesin, myosin, and dynein), \emph{E coli} chaperonin GroEL, $\mathrm{F_0F_1}$-ATPase, ribosomes, and helicases. 
A common theme in the function of these systems is they consume energy and in the process undergo a reaction cycle that dictates their function. 
Free energy transduction from chemical energy to mechanical work via a series of conformational switches is the hallmark of biological nanomachines \cite{Vale00Science}. 
\\

{\bf Chaperonin GroEL:}
Most of the proteins in cells fold spontaneously. 
However, molecular chaperones have evolved to rescue a small fraction of proteins, which do not reach their native states easily and hence are destined to aggregate. 
In \emph{E. coli} it is estimated that only about (5-10) \% of the proteins \cite{lorimer1996FASEBJ} require assistance from the chaperonin GroEL, which has been extensively characterized using experiments and simulations \cite{ThirumalaiARBBS01}. 

GroEL has two heptameric rings that are stacked back-to-back \cite{SiglerNature97}, with each subunit consisting of apical (A), intermediate (I), and equatorial (E) domains. 
During the reaction cycle GroEL (Fig. 5) undergoes a series of a structural (allosteric) transitions upon binding of SP, ATP, and the co-chaperonin GroES (Fig.\ref{GroEL}). 
In the $T$ state, the hydrophobic patches in the A-domain  recognize the exposed hydrophobic residues of the misfolded SPs.
ATP-binding triggers  dramatic domain movements in GroEl resulting in the catalytic sites moving apart, which in turn  imparts a stretching force  to partially unfold the captured SP. This step  is followed by GroES binding, which results in the encapsulation of the SP in the central cavity.  
 The extent of structural changes at the molecular level in each subunit of GroEL  (each ring has $\sim 3850$ residues) during the reaction cycle ($T\rightarrow R\rightarrow R"\rightarrow T$) was revealed only through CG simulations \cite{Hyeon06PNAS}.

Simulations using the SOP model of the entire heptameric GroEL particle vividly illustrated the conformational changes of GroEL triggered by ATP binding ($T \rightarrow  R$) and ATP hydrolysis ($R \rightarrow R"$). 
Multiple simulation trajectories revealed an unprecedented view of the key interactions that drive the allosteric
transitions \cite{Hyeon06PNAS}: 
(i) A domains rotate counterclockwise in the $T \rightarrow R$ transition and clockwise in $R \rightarrow R"$ transition. 
(ii) Global $T \rightarrow R$ and $R \rightarrow R"$ transitions follow two-state kinetics while the formation and kinetics of disruption of residue pairs encompass a broad range of time scales.    There is an underlying kinetic hierarchy of internal dynamics that govern global transitions. (iii) For both $T\rightarrow R$ and $R\rightarrow R"$ transitions, disruption and formation of salt-bridges are coordinated at multiple sites, which mediate the communication between two neighboring subunits and synchronize the dynamics of the heptameric ring. 
(iv) There is a spectacular outside-in movement of two helices  accompanied by interdomain salt-bridge formation, which are both solvent exposed in the $R$ state. As a result the microenvironment of the SP, which is predominantly hydrophobic in the $T$ state becomes progressively hydrophilic as the reaction cycle proceeds.

These large scale conformational changes are linked to function.
As long as misfolded proteins, which typically have exposed hydrophobic regions, are presented to GroEL they are captured. 
In the transitions ($T\rightarrow R$) and even more dramatically in $R\rightarrow R"$ the structural changes in the GroEL particle results in the interactions between SP and GroEL from being favorable in the $T$ state to unfavorable in the $R"$ state. 
Changes in the microenvironment results in the SP being placed in different part of the folding landscape from which it can fold with some probability during the life time of the $R$ and $R"$ states (Fig. 5). 
If the cycle is iterated multiple times, sufficient yield of the SP can be obtained as anticipated by the Iterative Annealing Mechanism (IAM) \cite{LorimerPNAS96,Tehver08JMB}. 
It is amusing to note that the mechanism of GroEL function is hauntingly similar to the simulated annealing protocol \cite{Kirkpatrick83Science} used in the context of NP hard problems.
Not surprisingly, nature has stumbled upon it apparently millions of years earlier. 
\\

{\bf Kinesins:}
Kinesins are  motors that transport cellular organelles along the network of cytoskeletal filaments \cite{Vale85Cell,Hirokawa98Science,Vale00Science}.
Made of two identical motor domains linked by a coiled-coil stalk, kinesins exploit the free energy generated from  binding and hydrolysis of ATP to produce the characteristic hand-over-hand stepping motion. 
A number of SM experiments show that kinesin takes roughly 8 nm step along the polar microtubule {\bf (MT)} track as it strides towards the $(+)$ end consuming one ATP per step. 

Due to the fundamental limitations in experimental resolution, it is difficult to provide molecular explanations of many intriguing observations related to kinesin motility such as the free energy transduction, out-of-phase coordination of the processes occurring at the two motor domains, and the role of kinesin-MT interactions.  
In order for both heads to associate with the MT binding sites, internal tension ($\sim (8-15)$ pN) exerted through the neck-linker deforms the catalytic site from its native-like configuration \cite{Hyeon07PNAS}, thus inhibiting the premature binding of ATP to the nucleotide free leading head.
The ATP inhibited state is maintained as long as the two heads remain bound. 
The deformed leading head catalytic site is restored only after the inorganic phosphate (P$_\mathrm{i}$) is released, which changes the trailing  head from a strong to a weak binding state.  
Thus, processivity of kinesin is regulated by strain in the leading head which can be linked to the topology of the kinesin-MT complex.  Simplified molecular simulations  combined with theoretical ideas have also shed light on the vexing question of whether kinesin takes substeps (Fig.\ref{kinesin}) \cite{Hyeon07PNAS2}. 
\\

{\bf Transcription initiation by bacterial RNA polymerase: }
The synthesis of RNA, carried out by DNA-dependent RNA polymerase (RNAP) in a process referred to as transcription, involves several stages. 
The highly regulated transcription process in eukaryotes is extraordinarily complicated involving a whole zoo of transcription factors which interact with polymerase as it reads the codes on  the template strand of DNA to make RNA (Fig.\ref{RNAP}). 
Transcription in bacteria also involves a number of steps. 
The DNA-dependent RNAP, whose sequence, structures, and global functions are universally conserved from bacteria to man, is the key enzyme in the transcription of genetic information in all organisms. The three major stages in the transcription cycle, which first involves binding of initiation-specific transcription factors to the catalytically competent core of RNAP, to form a holoenzyme are: (i) Initiation, during which initiation-specific $\sigma$ factor binds to the catalytically competent are RNAP to form the holoenzyme. 
This step is followed by recognition of the promoter DNA to form the closed ($R\cdot P_c$) complex and subsequent transition to the open ($R\cdot P_o$) structure. 
(ii) Elongation of the transcript by nucleotide addition. (iii) Termination involving cessation of transcription and disassembly of the RNAP elongation complex. 

Recently  the dynamics of structural transitions that occur during $R\cdot P_c\rightleftharpoons R\cdot P_o$ transition, which leads to melting of 12 base pairs in the promoter region resulting in the formation of transcription bubble (Fig.\ref{RNAP}b) \cite{Chen10PNAS} were probed using CG simulations \cite{Chen10PNAS}. 
To perform these simulations, CG model for the 3,122 residue RNAP-DNA complex (15 nm long and 11 nm wide) that is identical to those used to describe GroEL and kinesin dynamics, was used. 
For DNA, each strand was represented using a single site located at the center of the nucleotide. 
Transcription bubble forms in three steps (Fig.\ref{RNAP}c). 
(i) Melting of -10 element on the promoter region. 
(ii) Scrunching of promoter DNA into RNAP active channel, followed by the formation of bubble; the accommodation of dsDNA into the channel involves an internal RNAP dynamics of a transient expansion of key structural motifs in the $\beta$ subunit. 
(iii) Bending of downstream DNA after the unwinding of the dsDNA.
 Simulation results revealed that internal RNAP dynamics resulting in transient buildup of strain is needed to fully accommodate dsDNA to gain access to the active site. 
The simulations (for animation see http://www.youtube.com/watch?v=Q6QoyDl3TCw) also make several testable predictions to probe the relationship between RNAP motion and transcription bubble formation.  
\\
  
\section*{Outlook}

Given that biological problems are complex  it is inevitable that CG models should play a key role in informing experiments. Although not reviewed here there are a number of areas such as protein aggregation, membrane structure and dynamics \cite{Marrink07JPCB}, and lipid-membrane interactions where such simulations have already been profitable \cite{Straub10COSB}.  Experimental constraints and theory have been the guiding factors in constructing length-scale dependent CG models, as a few examples here illustrate. Current methods can be used to provide insights into a number of important biological problems.  
On a few nm length scale, corresponding to  folding problems, there is a need to study proteins in excess of 200 residues. 
Modeling counter ion effects to describe RNA folding can be achieved by integrating theoretical ideas from polymer physics and suitable CG models.   Although electrostatic interactions have been approximately modeled in simulations of biological machines \cite{Hyeon06PNAS} further refinements might needed for more accurate simulations \cite{Messer10Proteins}.   On longer length scales there are a number of problems which could profit from CG simulations.   
Description of motor-driven polymerization and depolymerization kinetics of microtubule and protein-induced polymerization of actin are two examples.

The demand to develop CG models will continue to grow because there is an appetite to understand  the workings of a cell. 
The increasing attention paid to obtain real-time measurements on how the workers (enzymes, ribozymes, ribosomes, genomes, lipids, membranes etc) cooperate to execute the demands on the cell is sure to spur interests in models and theories. 
From a modeling perspective, it is neither possible nor desirable to devise microscopic models when considering events on long length and time scales. 
In constructing whole-cell models, it may be sufficient to model the various workers as quasiparticles, which interact with each other through connected networks that are dynamically changing depending upon the cell status and external stimuli. 
Such a viewpoint is already being used in systems biology.  
The lesson from theoretical approaches to problems in condensed matter and material science is that phenomena at different length and time scales require different levels of description.  
Such a perspective, which also applies to biological problems, will surely spur us on to develop suitable CG models and theories that capture the essence of the problem at hand without being encumbered by unnecessary details.  \\

{\bf Acknowledgements:} This work was supported by a grant from Korea Research Foundation grants (2010-0000602 and KRF-C00180) (C.H.) and the National Science Foundation (CHE09-14033) (D.T.). The list of references is not exhaustive and should be viewed as a guide to the literature.
\\

{\bf Competing financial interest:} The authors declare no competing financial interests.
\clearpage

\renewcommand\figurename{Box.} 
\begin{wrapfigure}{r}{0.5\textwidth}
\includegraphics[width=3.5in]{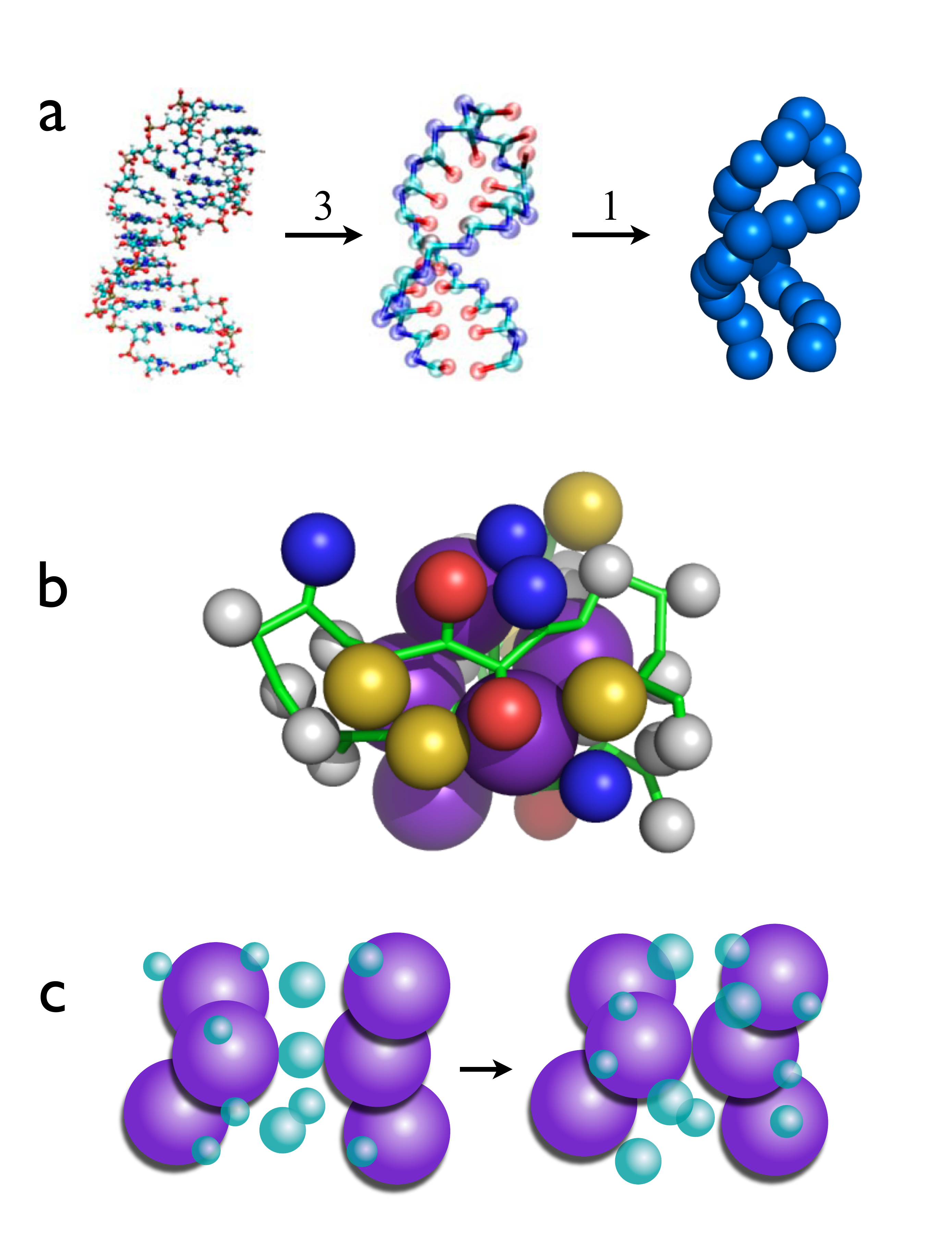}
\caption{
\label{Box_CG}}
\end{wrapfigure}

Box \ref{Box_CG}: {\bf Genre of CG models.}
Formally one can think of coarse-graining as a process by which effective energy functions are generated from microscopic Hamiltonian by integrating over irrelevant degrees of freedom. 
The block spin renormalization procedure in Ising model shows that as the degrees of freedom are thinned multiparticle interactions are generated. 
Similar ideas can be used to construct effective Hamiltonian by insisting that the partition function for the CG and the microscopic Hamiltonian be the same. However, given the large inaccuracies in force fields, intuitive and physical consideration have proved far more profitable in guiding the development of CG models. 
The CG strategy is successful because the characteristic time scales at each length scale are well separated from each other.

In response to the challenge of describing biological processes that span several orders of magnitude in time and length scales a variety of CG models for DNA, RNA and proteins have been proposed. 
Although CG models have been prevalent in the polymer literature for over fifty years their use in proteins began in earnest with the pioneering work of Levitt and Warshel \cite{Levitt75Nature}.  The efficacy of off-lattice models for protein folding kinetics was first demonstrated by Honeycutt and Thirumalai \cite{Honeycutt90PNAS}.  
In all the CG models  polypeptide chains and nucleic acids are represented using a reduced description.  The accompanying figure shows a few examples. 
{\bf a}. Three Interaction Site (TIS) models for a RNA hairpin obtained by representing each nucleotide by three sites one each for phosphate, ribose and base. 
{\bf b}. $C_{\alpha}$-SCM for WW domain obtained by replacing each amino acid by two sites one centered on the $\alpha$-carbon and the other at the center of the side chain. 
{\bf c}. Water-mediated interactions can also be captured using effective potentials \cite{papoian2003JACS}. 
Using these representations and their variations a number of types of CG models have been developed. The common unifying aspect of all these models is that the nucleotides and amino acids are represented by only a few interaction sites. However, they vary in details especially the number of interaction centers per nucleotide or amino acid needed to encode the folded structure. 

A major advantage of CG models is that their conformational space can be exhaustively sampled. 
However, even with simplification accurate results for thermodynamics can only be obtained using enhanced sampling methods. 
Towards this end simulation of CG models have used replica exchange methods and multicanonical methods. In addition, low friction Langevin dynamics has also been used to efficiently sample conformational space. These methods are necessary especially in simulating proteins with complex topology.  In order to obtain kinetic information for folding or transition times between allosteric states typically Brownian dynamics (BD) simulations are performed. 
In typical BD simulations the Brownian time is $\tau_H\approx \zeta_Ha^2/k_BT_s$ where $\zeta_H$ is the friction constant, $a$ is the roughly the size of a coarse-grained bead, and $T_s$ is the simulation temperature. 
Estimate of these quantities \cite{VeitshansFoldDes97} have been used to map simulation times to real times in a number of applications. \\


\begin{thebibliography}{10}

\bibitem{Anderson:1997}
Anderson, P.~W.
\newblock (1997) {\em {Basic Notions of Condensed Matter Physics}}.
\newblock (Westview Press).\\
\newblock {{\bf ** This classic monograph describes powerful strategies for tackling the many-body problem by using effective Hamiltonians.}}

\bibitem{Mezardbook}
M{\'e}zard, M, Parisi, G,  \& Virasoro, M.
\newblock (1988) {\em Spin glass theory and beyond}.
\newblock (World Scientific).

\bibitem{Kirk89JPhysA}
Kirkpatrick, T \& Thirumalai, D.
\newblock {Random solution from regular density functional Hamiltonian - A
  static and dynamical theory for the structural glass transition}.
\newblock {\em {J. Phys. A. }} {\bf {22}}, {L149--L155}.
\newblock ({1989}).

\bibitem{Vas07ARPC}
Lubchenko, V \& Wolynes, P.
\newblock {Theory of structural glasses and supercooled liquids}.
\newblock {\em {Ann. Rev. Phys. Chem.}} {\bf {58}}, {235--266}.
\newblock ({2007}).

\bibitem{Alder57JCP}
Alder, B.~J \& Wainwright, T.~E.
\newblock {Phase transition for a hard sphere system}.
\newblock {\em J. Chem. Phys.} {\bf 27}, 1208.
\newblock (1957).

\bibitem{Hyeon05BC}
Thirumalai, D \& Hyeon, C.
\newblock {RNA and Protein folding: Common Themes and Variations}.
\newblock {\em Biochemistry} {\bf 44}, 4957--4970.
\newblock (2005).

\bibitem{OnuchicCOSB04}
Onuchic, J.~N \& Wolynes, P.~G.
\newblock Theory of protein folding.
\newblock {\em Curr. Opin. Struct. Biol.} {\bf 14}, 70--75.
\newblock (2004).

\bibitem{Shakhnovich06Chemrev}
Shakhnovich, E.
\newblock Protein folding thermodynamics and dynamics: Where physics,
  chemistry, and biology meet.
\newblock {\em Chem. Rev.} {\bf 106}, 1559--1588.
\newblock (2006).

\bibitem{Dill08ARB}
Dill, K.~A, Ozkan, S.~B, Shell, M.~S,  \& Weikl, T.~R.
\newblock The protein folding problem.
\newblock {\em Annu. Rev. Biophys.} {\bf 37}, 289--316.
\newblock (2008).

\bibitem{Thirumalai10ARB}
Thirumalai, D, {O'Brien}, E.~P, Morrison, G,  \& Hyeon, C.
\newblock Theoretical perspectives on protein folding.
\newblock {\em Annu. Rev. Biophys.} {\bf 39}, 159--183.
\newblock (2010).

\bibitem{Levitt75Nature}
Levitt, M \& Warshel, A.
\newblock Computer simulation of protein folding.
\newblock {\em Nature} {\bf 253}, 694--698.
\newblock (1975).

\bibitem{Honeycutt90PNAS}
Honeycutt, J.~D \& Thirumalai, D.
\newblock Metastability of the folded states of globular proteins.
\newblock {\em Proc. Natl. Acad. Sci. USA} {\bf 87}, 3526--3529.
\newblock (1990).

\bibitem{Clementi00JMB}
Clementi, C, Nymeyer, H,  \& Onuchic, J.~N.
\newblock Topological and energetic factors: what determines the structural
  details of the transition state ensemble and ``en-route" intermediates for
  protein folding? an investigation for small globular protein.
\newblock {\em J. Mol. Biol.} {\bf 298}, 937--953.
\newblock (2000).

\bibitem{Karanicolas2002ProtSci}
Karanicolas, J \& Brooks~III, C.
\newblock {The origins of asymmetry in the folding transition states of protein
  L and protein G}.
\newblock {\em Protein Science} {\bf 11}, 2351--2361.
\newblock (2002).

\bibitem{Weinkam10ACR}
Weinkam, P, Zimmermann, J, Romesberg, F,  \& Wolynes, P.
\newblock The folding energy landscape and free energy excitations of
  cytochrome c.
\newblock {\em Acc. Chem. Res.} {\bf 43}, 652--660.
\newblock (2010).

\bibitem{Hyeon05PNAS}
Hyeon, C \& Thirumalai, D.
\newblock {Mechanical unfolding of RNA hairpins}.
\newblock {\em Proc. Natl. Acad. Sci. USA} {\bf 102}, 6789--6794.
\newblock (2005).

\bibitem{Whitford10RNA}
Whitford, P.~C, Geggler, P, Altman, R.~B, Blanchard, S.~C, Onuchic, J.~N,  \&
  Sanbonmatsu, K.~Y.
\newblock {Accommodation of aminoacyl-tRNA into the ribosome involves
  reversible excursions along multiple pathways}.
\newblock {\em RNA} {\bf 16}, 1196--1204.
\newblock (2010).

\bibitem{Hyeon06PNAS}
Hyeon, C, Lorimer, G.~H,  \& Thirumalai, D.
\newblock {Dynamics of allosteric transition in GroEL}.
\newblock {\em Proc. Natl. Acad. Sci. USA} {\bf 103}, 18939--18944.
\newblock (2006).\\
\newblock {{\bf * Coarse-grained SOP model was used to study the molecular details in the conformational changes that occur during the reaction cycle of the annealing machine GroEL.}}

\bibitem{Pisliakov09PNAS}
Pisliakov, A, Cao, J, Kamerlin, S,  \& Warshel, A.
\newblock Enzyme millisecond conformational dynamics do not catalyze the
  chemical step.
\newblock {\em Proc. Natl. Acad. Sci. USA} {\bf 106}, 17359.
\newblock (2009).

\bibitem{Rychkova2010PNAS}
Rychkova, A, Vicatos, S,  \& Warshel, A.
\newblock On the energetics of translocon-assisted insertion of charged
  transmembrane helices into membranes.
\newblock {\em Proc. Natl. Acad. Sci. USA} {\bf 107},
  17598.
\newblock (2010).

\bibitem{Strajbl2003PNAS}
Strajbl, M, Shurki, A,  \& Warshel, A.
\newblock Converting conformational changes to electrostatic energy in
  molecular motors: The energetics of ATP synthase.
\newblock {\em Proc. Natl. Acad. Sci. USA} {\bf 100}, 14834.
\newblock (2003).

\bibitem{Koga06PNAS}
Koga, N \& Takada, S.
\newblock Folding-based molecular simulations reveal mechanisms of the rotary
  motor $f_1$-atpase.
\newblock {\em Proc. Natl. Acad. Sci. USA} {\bf 103}, 5367--5372.
\newblock (2006).

\bibitem{Hyeon07PNAS}
Hyeon, C \& Onuchic, J.~N.
\newblock Internal strain regulates the nucleotide binding site of the kinesin
  leading head.
\newblock {\em Proc. Natl. Acad. Sci. USA} {\bf 104}, 2175--2180.
\newblock (2007).

\bibitem{Hyeon07PNAS2}
Hyeon, C \& Onuchic, J.~N.
\newblock Mechanical control of the directional stepping dynamics of the
  kinesin motor.
\newblock {\em Proc. Natl. Acad. Sci. USA} {\bf 104}, 17382--17387.
\newblock (2007).\\
\newblock {{\bf * Employing multi-scale CG simulations, this work studied the complex stepping dynamics resulting from the interplay between neck-linker docking and diffusive search dynamics of a kinesin head on the microtubule surface. }}


\bibitem{Takano10PNAS}
Takano, M, Terada, T,  \& Sasai, M.
\newblock {Unidirectional Brownian motion observed in an in silico single
  molecule experiment of an actomyosin motor}.
\newblock {\em Proc. Natl. Acad. Sci. USA} {\bf 107}, 7769.
\newblock (2010).

\bibitem{Tehver10Structure}
Tehver, R \& Thirumalai, D.
\newblock {Rigor to Post-Rigor Transition in Myosin V: Link between the
  Dynamics and the Supporting Architecture}.
\newblock {\em Structure} {\bf 18}, 471--481.
\newblock (2010).

\bibitem{Liu09PNAS}
Liu, H, Shi, Y, Chen, X,  \& Warshel, A.
\newblock Simulating the electrostatic guidance of the vectorial translocations
  in hexameric helicases and translocases.
\newblock {\em Proc. Natl. Acad. Sci. USA} {\bf 106},
  7449.
\newblock (2009).

\bibitem{Kamerlin11ARPC}
Kamerlin, S, Vicatos, S, Dryga, A,  \& Warshel, A.
\newblock Coarse-grained (multiscale) simulations in studies of biophysical and
  chemical systems.
\newblock {\em Annu. Rev. Phys. Chem.} {\bf 62}, 41--64.
\newblock (2011).

\bibitem{deGennesbook}
{de Gennes}, P.~G.
\newblock (1979) {\em {Scaling Concepts in Polymer Physics}}.
\newblock (Cornell University Press, Ithaca and London).

\bibitem{Minton08ARB}
Zhou, H.~X, Rivas, G,  \& Minton, A.~P.
\newblock Macromolecular crowding and confinement: Biochemical, biophysical,
  and potential physiological consequences.
\newblock {\em Ann. Rev. Biophys.} {\bf 37}, 375--397.
\newblock (2008).

\bibitem{Elcock10COSB}
Elcock, A.
\newblock {Models of macromolecular crowding effects and the need for
  quantitative comparisons with experiment}.
\newblock {\em Curr. Opin. Struct. Biol.} {\bf 20}, 196--206.
\newblock (2010).

\bibitem{Cheung05PNAS}
Cheung, M.~S, Klimov, D,  \& Thirumalai, D.
\newblock Molecular crowding enhances native state stability and refolding
  rates of globular proteins.
\newblock {\em Proc. Natl. Acad. Sci. USA} {\bf 102}, 4753--4758.
\newblock (2005).

\bibitem{Fisher74RMP}
Fisher, M.~E.
\newblock {Renormalization group in theory of critical behavior}.
\newblock {\em {Rev. Mod. Phys.}} {\bf {46}}, {597--616}.
\newblock ({1974}).

\bibitem{Bustamante94SCI}
Bustamante, C, Marko, J.~F, Siggia, E.~D,  \& Smith, S.
\newblock Entropic elasticity of $\lambda$-phase {DNA}.
\newblock {\em Science} {\bf 265}, 1599--1600.
\newblock (1994).

\bibitem{VallePRL05}
Valle, F, Favre, M, {de Los Rios}, P, Rosa, A,  \& Dietler, G.
\newblock Scaling exponents and probability distribution of DNA end-to-end
  distance.
\newblock {\em Phys. Rev. Lett.} {\bf 95}, 158105.
\newblock (2005).

\bibitem{PastorJCP96}
Pastor, R.~W, Zwanzig, R,  \& Szabo, A.
\newblock Diffusion limited first contact of the ends of a polymer: Comparison
  of theory with simulation.
\newblock {\em J. Chem. Phys.} {\bf 105}, 3878--3882.
\newblock (1996).

\bibitem{Toan08JPCB}
Toan, N, Greg~Morrison, P, Hyeon, C, Thirumalai, D,  et~al.
\newblock {Kinetics of Loop Formation in Polymer Chains }.
\newblock {\em J. Phys. Chem. B} {\bf 112}, 6094--6106.
\newblock (2008).

\bibitem{VologodskiiMacro00}
Podtelezhnikov, A.~A \& Vologodskii, A.~V.
\newblock {Dynamics of Small Loops in DNA Molecules}.
\newblock {\em Macromolecules} {\bf 33}, 2767--2771.
\newblock (2000).

\bibitem{Vologodskii92JMB}
Vologodskii, A, Levene, S, Klenin, K, Frank-Kamenetskii, M,  \& Cozzarelli, N.
\newblock {Conformational and thermodynamic properties of supercoiled DNA}.
\newblock {\em J. Mol. Biol.} {\bf 227}, 1224--1243.
\newblock (1992).

\bibitem{VologodskiiMacro97}
Podtelezhnikov, A \& Vologodskii, A.
\newblock {Simulations of Polymer Cyclization by Brownian Dynamics}.
\newblock {\em Macromolecules} {\bf 30}, 6668--6673.
\newblock (1997).

\bibitem{HaELett03}
Jun, S, Bechhoefer, J,  \& Ha, B.-Y.
\newblock Diffusion-limited loop formation of semiflexible polymers: Kramers
  theory and the interwined time scales of chain relaxation and closing.
\newblock {\em Europhys. Lett.} {\bf 64}, 420--426.
\newblock (2003).

\bibitem{HyeonJCP06}
Hyeon, C \& Thirumalai, D.
\newblock Kinetics of interior loop formation in semiflexible chains.
\newblock {\em J. Chem. Phys.} {\bf 124}, 104905.
\newblock (2006).

\bibitem{WidomMC04}
Cloutier, T \& Widom, J.
\newblock {Spontaneous Sharp Bending of Double-Stranded DNA}.
\newblock {\em Mol. Cell.} {\bf 14}, 355--362.
\newblock (2004).

\bibitem{VologodskiiPNAS05}
Du, Q, Smith, C, Shiffeldrim, N, Vologodskaia, M,  \& Vologodskii, A.
\newblock {Cyclization of short DNA fragments and bending fluctuations of the
  double helix}.
\newblock {\em Proc. Natl. Acad. Sci. USA} {\bf 102}, 5397--5402.
\newblock (2005).

\bibitem{Savelyev10PNAS}
Savelyev, A \& Papoian, G.
\newblock {Chemically accurate coarse graining of double-stranded DNA}.
\newblock {\em Proc. Natl. Acad. Sci. USA}.
\newblock (2010).

\bibitem{BustamanteSCI92}
Smith, S.~B, Finzi, L,  \& Bustamante, C.
\newblock {Direct Mechanical Measurements of the Elasticity of Single DNA
  Molecules by Using Magnetic Beads}.
\newblock {\em Science} {\bf 258}, 1122--1126.
\newblock (1992).

\bibitem{jun06PNAS}
Jun, S \& Mulder, B.
\newblock {Entropy-driven spatial organization of highly confined polymers:
  lessons for the bacterial chromosome}.
\newblock {\em Proc. Natl. Acad. Sci. USA} {\bf 103}, 12388.
\newblock (2006).\\
\newblock {{\bf* Using CG models and polymer physics concepts an entropy-driven mechanism for DNA segregation in bacterial cells is proposed.}}

\bibitem{jun2010NRM}
Jun, S \& Wright, A.
\newblock {Entropy as the driver of chromosome segregation}.
\newblock {\em Nature Rev. Microbiol.} {\bf 8}, 600--607.
\newblock (2010).

\bibitem{cremer2001NRG}
Cremer, T \& Cremer, C.
\newblock {Chromosome territories, nuclear architecture and gene regulation in
  mammalian cells}.
\newblock {\em Nature Rev. Genet.} {\bf 2}, 292--301.
\newblock (2001).

\bibitem{grosberg1993EPL}
Grosberg, A, Rabin, Y, Havlin, S,  \& Neer, A.
\newblock {Crumpled globule model of the three-dimensional structure of DNA}.
\newblock {\em Europhys. Lett.} {\bf 23}, 373.
\newblock (1993).

\bibitem{lieberman09Science}
Lieberman-Aiden, E, van Berkum, N, Williams, L, Imakaev, M, Ragoczy, T,
  Telling, A, Amit, I, Lajoie, B, Sabo, P, Dorschner, M,  et~al.
\newblock {Comprehensive mapping of long-range interactions reveals folding
  principles of the human genome}.
\newblock {\em Science} {\bf 326}, 289.
\newblock (2009).\\
\newblock {{\bf * A polymer model that represents $\sim$1Mb DNA as a CG interaction center was used to explain the interphase DNA organization of the human genome, which is postulated to be fractal globule.}}


\bibitem{grosberg1988JP}
Grosberg, A, Nechaev, S,  \& Shakhnovich, E.
\newblock {The role of topological constraints in the kinetics of collapse of
  macromolecules}.
\newblock {\em J. Phys.} {\bf 49}, 2095--2100.
\newblock (1988).

\bibitem{ThirumARPC01}
Thirumalai, D, Lee, N, Woodson, S.~A,  \& Klimov, D.~K.
\newblock {Early Events in RNA Folding}.
\newblock {\em Annu. Rev. Phys. Chem.} {\bf 52}, 751--762.
\newblock (2001).

\bibitem{DimaJMB05}
Dima, R.~I, Hyeon, C,  \& Thirumalai, D.
\newblock Extracting stacking interaction parameters for {RNA} from the data
  set of native structures.
\newblock {\em J. Mol. Biol.} {\bf 347}, 53--69.
\newblock (2005).

\bibitem{moghaddam09JMB}
Moghaddam, S, Caliskan, G, Chauhan, S, Hyeon, C, Briber, R, Thirumalai, D,  \&
  Woodson, S.
\newblock {Metal ion dependence of cooperative collapse transitions in RNA}.
\newblock {\em J. Mol. Biol.} {\bf 393}, 753--764.
\newblock (2009).

\bibitem{Koculi07JACS}
Koculi, E, Hyeon, C, Thirumalai, D,  \& Woodson, S.~A.
\newblock {Charge Density of Divalent Metal Cations Determines RNA Stability}.
\newblock {\em J. Am. Chem. Soc.} {\bf 129}, 2676--2682.
\newblock (2007).

\bibitem{RussellPNAS02}
Russell, R, Millett, I.~S, Tate, M.~W, Kwok, L.~W, Nakatani, B, Gruner, S.~M,
  Mochrie, S. G.~J, Pande, V, Doniach, S, Herschlag, D,  \& Pollack, L.
\newblock Rapid compaction during {RNA} folding.
\newblock {\em Proc. Natl. Acad. Sci. USA} {\bf 99}, 4266--4271.
\newblock (2002).\\
\newblock {{\bf * This work used a CG model to infer the structures of {\em T.} ribozyme at each stage of collapse dynamics by selecting a conformation that best explain the SAXS profile}}

\bibitem{Ma06JACS}
Ma, H, Proctor, D.~J, Kierzek, E, Kierzek, R, Bevilacqua, P.~C,  \& Gruebele,
  M.
\newblock {Exploring the energy landscape of a small RNA hairpin}.
\newblock {\em J. Am. Chem. Soc.} {\bf 128}, 1523--1530.
\newblock (2006).

\bibitem{Hyeon08JACS}
Hyeon, C \& Thirumalai, D.
\newblock {Multiple probes are required to explore and control the rugged
  energy landscape of RNA hairpins}.
\newblock {\em J. Am. Chem. Soc.} {\bf 130}, 1538--1539.
\newblock (2008).

\bibitem{Chen00PNAS}
Chen, S.~J \& Dill, K.~A.
\newblock {RNA folding energy landscapes}.
\newblock {\em Proc. Natl. Acad. Sci. USA} {\bf 97}, 646--651.
\newblock (2000).

\bibitem{ThirumCOSB99}
Thirumalai, D \& Klimov, D.~K.
\newblock {Deciphering the Time Scales and Mechanisms of Protein Folding Using
  Minimal Off-Lattice Models}.
\newblock {\em Curr. Opin. Struct. Biol.} {\bf 9}, 197--207.
\newblock (1999).

\bibitem{Obrien08PNAS}
{O'Brien}, E, Ziv, G, Haran, G, Brooks, B,  \& Thirumalai, D.
\newblock {Effects of denaturants and osmolytes on proteins are accurately
  predicted by the molecular transfer model}.
\newblock {\em Proc. Natl. Acad. Sci. USA} {\bf 105}, 13403.
\newblock (2008).

\bibitem{liu2011PNAS}
Liu, Z, Reddy, G, OÕBrien, E,  \& Thirumalai, D.
\newblock Collapse kinetics and chevron plots from simulations of
  denaturant-dependent folding of globular proteins.
\newblock {\em Proc. Natl. Acad. Sci. USA} {\bf 108},
  7787.
\newblock (2011).
\newblock {{\bf * The first demonstration that coarse-grained models can be used to characterize the folding thermodynamics and kinetics of a globular protein in the presence of denaturants.}}

\bibitem{tinoco2006QRB}
Tinoco, I, TX~Li, P,  \& Bustamante, C.
\newblock {Determination of thermodynamics and kinetics of RNA reactions by
  force}.
\newblock {\em Q. Rev. Biophys.} {\bf 39}, 325--360.
\newblock (2006).

\bibitem{Fernandez04Science}
Fernandez, J.~M \& Li, H.
\newblock Force-clamp spectroscopy monitors the folding trajectory of a single
  protein.
\newblock {\em Science} {\bf 303}, 1674--1678.
\newblock (2004).

\bibitem{Hyeon08PNAS}
Hyeon, C, Morrison, G,  \& Thirumalai, D.
\newblock {Force dependent hopping rates of RNA hairpins can be estimated from
  accurate measurement of the folding landscapes}.
\newblock {\em Proc. Natl. Acad. Sci. USA} {\bf 105}, 9604--9606.
\newblock (2008).

\bibitem{HyeonMorrison09PNAS}
Hyeon, C, Morrison, G, Pincus, D.~L,  \& Thirumalai, D.
\newblock Refolding dynamics of stretched biopolymers upon force-quench.
\newblock {\em Proc. Natl. Acad. Sci. USA} {\bf 106}, 20288--20293.
\newblock (2009).

\bibitem{Dudko06PRL}
Dudko, O.~K, Hummer, G,  \& Szabo, A.
\newblock Intrinsic rates and activation free energies from single-molecule
  pulling experiments.
\newblock {\em Phys. Rev. Lett.} {\bf 96}, 108101.
\newblock (2006).

\bibitem{Hyeon03PNAS}
Hyeon, C \& Thirumalai, D.
\newblock Can energy landscape roughness of proteins and {RNA} be measured by
  using mechanical unfolding experiments?
\newblock {\em Proc. Natl. Acad. Sci. USA} {\bf 100}, 10249--10253.
\newblock (2003).

\bibitem{Hyeon06Structure}
Hyeon, C, Dima, R.~I,  \& Thirumalai, D.
\newblock {Pathways and kinetic barriers in mechanical unfolding and refolding
  of RNA and proteins}.
\newblock {\em Structure} {\bf 14}, 1633--1645.
\newblock (2006).

\bibitem{Mickler07PNAS}
Mickler, M, Dima, R.~I, Dietz, H, Hyeon, C, Thirumalai, D,  \& Rief, M.
\newblock {Revealing the bifurcation in the unfolding pathways of GFP by using
  single-molecule experiments and simulations}.
\newblock {\em Proc. Natl. Acad. Sci. USA} {\bf 104}, 20268--20273.
\newblock (2007).

\bibitem{YonathCell01}
Harms, J, Schluenzen, F, Zarivach, R, Bashan, A, Gat, S, Agmon, I, Bartels, H,
  Franceschi, F,  \& Yonath, A.
\newblock High resolution structure of the large ribosomal subunit from a
  mesophilic eubacterium.
\newblock {\em Cell} {\bf 107}, 679--688.
\newblock (2001).

\bibitem{SteitzARB03}
Moore, P.~B \& Steitz, T.~A.
\newblock The structural basis of large ribosomal subunit function.
\newblock {\em Ann. Rev. Biochem} {\bf 72}, 813--850.
\newblock (2003).

\bibitem{Ziv2005PNAS}
Ziv, G, Haran, G,  \& Thirumalai, D.
\newblock {Ribosome exit tunnel can entropically stabilize $\alpha$-helices}.
\newblock {\em Proc. Natl. Acad. Sci. USA} {\bf 102}, 18956.
\newblock (2005).

\bibitem{Elcock2006PLOSCompBio}
Elcock, A.
\newblock {Molecular simulations of cotranslational protein folding: fragment
  stabilities, folding cooperativity, and trapping in the ribosome}.
\newblock {\em PLoS Comput. Biol} {\bf 2}, 824--41.
\newblock (2006).

\bibitem{obrien10JACS}
{OÕBrien}, E.~P, Hsu, S. T.~D, Christodoulou, J, Vendruscolo, M,  \& Dobson,
  C.~M.
\newblock {Transient Tertiary Structure Formation within the Ribosome Exit
  Port}.
\newblock {\em J. Am. Chem. Soc.} {\bf 132}, 16928--16937.
\newblock (2010).

\bibitem{phillips2009physical}
Phillips, R, Kondev, J, Theriot, J, Orme, N,  \& Garcia, H.
\newblock {Physical biology of the cell}.
\newblock (2009).

\bibitem{minton2001JBC}
Minton, A.
\newblock {The influence of macromolecular crowding and macromolecular
  confinement on biochemical reactions in physiological media}.
\newblock {\em J. Biol. Chem.} {\bf 276}, 10577.
\newblock (2001).

\bibitem{dhar10PNAS}
Dhar, A, Samiotakis, A, Ebbinghaus, S, Nienhaus, L, Homouz, D, Gruebele, M,  \&
  Cheung, M.
\newblock {Structure, function, and folding of phosphoglycerate kinase are
  strongly perturbed by macromolecular crowding}.
\newblock {\em Proc. Natl. Acad. Sci. USA} {\bf 107}, 17586--17591.
\newblock (2010).\\
\newblock {{\bf * This work shows that the presence of crowders is critical in shaping the functional state of phosphoglycerate kinase.}}

\bibitem{Vale00Science}
Vale, R.~D \& Milligan, R.~A.
\newblock The way things move: Looking under the hood of molecular motor
  proteins.
\newblock {\em Science} {\bf 288}, 88--95.
\newblock (2000).

\bibitem{lorimer1996FASEBJ}
Lorimer, G.
\newblock {A quantitative assessment of the role of the chaperonin proteins in
  protein folding in vivo}.
\newblock {\em The FASEB Journal} {\bf 10}, 5.
\newblock (1996).

\bibitem{ThirumalaiARBBS01}
Thirumalai, D \& Lorimer, G.~H.
\newblock Chaperonin-mediated protein folding.
\newblock {\em Ann. Rev. Biophys. Biomol. Struct.} {\bf 30}, 245--269.
\newblock (2001).

\bibitem{SiglerNature97}
Xu, Z, Horwich, A.~L,  \& Sigler, P.~B.
\newblock The crystal structure of the asymmetric {GroEL-GroES-(ADP)$_7$}
  chaperonin complex.
\newblock {\em Nature} {\bf 388}, 741.
\newblock (1997).

\bibitem{LorimerPNAS96}
Todd, M.~J, Lorimer, G.~H,  \& Thirumalai, D.
\newblock Chaperonin-facilitated protein folding: Optimization of rate and
  yield by an iterative annealing mechanism.
\newblock {\em Proc. Natl. Acad. Sci. USA} {\bf 93}, 4030--4035.
\newblock (1996).

\bibitem{Tehver08JMB}
Tehver, R \& Thirumalai, D.
\newblock Kinetic model for the coupling between allosteric transitions in
  groel and substrate protein folding and aggregation.
\newblock {\em J. Mol. Biol.} {\bf 4}, 1279--1295.
\newblock (2008).

\bibitem{Kirkpatrick83Science}
Kirkpatrick, S, Gelatt, D,  \& Vecchi, M.~P.
\newblock Optimization by simulated annealing.
\newblock {\em Science} {\bf 220}, 671--680.
\newblock (1983).

\bibitem{Vale85Cell}
Vale, R.~D, Reese, T.~S,  \& Sheetz, M.~P.
\newblock Identification of a novel force-generating protein, kinesin, involved
  in microtubule-based motility.
\newblock {\em Cell} {\bf 42}, 39--50.
\newblock (1985).

\bibitem{Hirokawa98Science}
Hirokawa, N.
\newblock Kinesin and dynein superfamily proteins and the mechanism of
  organelle transport.
\newblock {\em Science} {\bf 279}, 519--526.
\newblock (1998).

\bibitem{Chen10PNAS}
Chen, J, Darst, S.~A,  \& Thirumalai, D.
\newblock {Promoter melting triggered by bacterial RNA polymerase occurs in
  three steps}.
\newblock {\em Proc. Natl. Acad. Sci. USA} {\bf 107}, 12523--12528.
\newblock (2010).\\
\newblock {{\bf * SOP model of RNA polymerase and dsDNA was used to reveal the molecular mechanism of melting of promoter region upon polymerase binding, which is is the first step in transcription initiation. }}

\bibitem{Marrink07JPCB}
{Marrink, S.J. and Risselada, H.J. and Yefimov, S. and Tieleman, D.P. and De
  Vries, A.H.}
\newblock {The MARTINI Force Field: Coarse Grained Model for Biomolecular
  Simulations}.
\newblock {\em {J. Phys. Chem. B}} {\bf 111}, 7812Ð7824.
\newblock (2007).

\bibitem{Straub10COSB}
Straub, J \& Thirumalai, D.
\newblock {Principles governing oligomer formation in amyloidogenic peptides}.
\newblock {\em Curr. Opin. Struct. Biol.} {\bf 20}, 187--195.
\newblock (2010).

\bibitem{Messer10Proteins}
Messer, B, Roca, M, Chu, Z, Vicatos, S, Kilshtain, A,  \& Warshel, A.
\newblock Multiscale simulations of protein landscapes: Using coarse-grained
  models as reference potentials to full explicit models.
\newblock {\em Proteins: Struct. Func. Bioinfo.} {\bf 78},
  1212--1227.
\newblock (2010).
\newblock {{\bf * The study demonstrates how different simulation strategies can be used to obtain insights into dtnamics of complex biological systems. }}

\bibitem{papoian2003JACS}
Papoian, G, Ulander, J,  \& Wolynes, P.
\newblock {Role of Water Mediated Interactions in Protein- Protein Recognition
  Landscapes}.
\newblock {\em J. Am. Chem. Soc} {\bf 125}, 9170--9178.
\newblock (2003).

\bibitem{VeitshansFoldDes97}
Veitshans, T, Klimov, D,  \& Thirumalai, D.
\newblock Protein folding kinetics: timescales, pathways and energy landscapes
  in terms of sequence-dependent properties.
\newblock {\em Folding Des.} {\bf 2}, 1--22.
\newblock (1997).

\end{thebibliography}

\clearpage 

\section*{Figure Captions}
Fig.\ref{DNA}: {\bf DNA applications.}
{\bf a.} Loop formation times between two regions in dsDNA separated by $s$ along the contour from simulations using CG model that represents a single-pitch of DNA helix as a monomer unit. Lines are theoretical results. {\bf b.} Extension as a function of mechanical force for 97kb $\lambda$-DNA. Symbols are experimental results and the dashed line is the fit using WLC model.  {\bf c.} Model of bacterial chromosomal separation from simulations of tightly confined polymer chain. The newly synthesized DNA (blue and red) is extruded to the periphery of the  unreplicated nucleoid (grey) and the two strings of blobs drift apart and segregate due to the excluded-volume interactions and
conformational entropy. {\bf d.} Top figure shows  scaling law  $P(s)\sim s^{-1.08}$ where $P(s)$ is the contact probability for a given genomic distance $s$, measured by Hi-C, a method that probes the three-dimensional architecture of whole genomes by coupling proximity-based ligation with massively parallel sequencing.  The exponent in the power law decay is  distinct from $s^{-1.5}$ for an  equilibrated globule (bottom left) whereas $s^{-1.08}$ scaling (dashed lines from CG simulations in the top figure) is explained using a fractal globule (bottom right), a knot-free, polymer conformation, which enables reversible folding and unfolding at any genomic locus. 
Figures {\bf a}$-${\bf d} were adapted from \cite{HyeonJCP06}, \cite{Bustamante94SCI}, \cite{jun2010NRM}, and \cite{lieberman09Science}, respectively. 
\\

Fig.\ref{RNA}: {\bf Ribozyme to RNA hairpin folding.}
{\bf a.} Left is the secondary structure map of  {\it Tetrahymena} group I intron where the circles show that between 5-6 nucleotides are used to represent one interaction center. Simulations of the CG model are used to obtain best agreement with time dependent SAXS signals as the ribozyme folds. Representative structures that produce best agreement with experiments are shown \cite{RussellPNAS02}.
{\bf b.} Refolding pathways of a RNA hairpin upon quenching the force from a high to low value (left) and obtained from temperature quench (right) using SOP model. 
Upon force quench folding commences from an extended ($E$) state by forming the turn, which nucleated the hairpin formation (left). However, folding occurs by multiple pathways upon temperature quench. 
\\

Fig.\ref{ProteinsFig}: {\bf Protein folding.}
{\bf a.} Dependence of fraction of molecules in the Native Basin of Attraction as function of Guanidinium Chloride concentration for protein L (blue) and Cold Shock protein (red) where the symbols are data from experiments, and the lines are results from Molecular Transfer Model simulations. 
{\bf b.} Unfolding of GFP, a $\sim$ 250-residue 11-stranded $\beta$-barrel protein (left). Forced unfolding obtained from CG simulations on the right shows pathway bifurcation. The structures from the simulations at various stages of unfolding are also shown.
{\bf c.} The volume of exit tunnel (left) and a helix (right) are shown in the ribosome structure.  
Cotranslational folding of a protein occurs as it is synthesized by the ribosome.  
\\

Fig.\ref{crowding}: {\bf Crowding effects on folding.} 
{\bf a.} Crowding-induced entropic stabilization of the folded states of proteins. Restriction of the extended denatured state ensemble because of volume occupied by crowding raises its free energy to a greater than the folded state. The structures are from $C_{\alpha}$-SCM simulations of a three stranded $\beta$-sheet protein, WW domain.
{\bf b.} Folding time as a function of the concentration of the crowding agent (black lines) for phosphoglycerate kinase (PGK). The red curve shows folding of WW domain as a function of $\phi_c$. The comparison is meant to illustrate that CG simulations qualitatively explain the measurements.
{\bf c.} Structure of PGK in the absence of crowding agent (left) and at $\phi_c = 0.25$ on the right. The distance between the N and C terminus lobes has been dramatically reduced by crowding. The functional implications are given in the text and in \cite{dhar10PNAS}.
Figures {\bf a}$-${\bf c} were adapted from \cite{Cheung05PNAS} and \cite{dhar10PNAS}. 
 \\

Fig.\ref{GroEL}: {\bf Reaction cycle and GroEL function.}
The hemicycle of GroEL reaction cycle, which shows that a misfolded substrate protein (SP) is captured by GroEL ({\it E. Coli.} chaperonin) in the $T$ state. This step is followed by reversible ATP-driven transition to the $R$ state to which the co-chaperonin GroES can bind to form the $R^{\prime}$ complex, which also results in the SP being encapsulated in the GroEL cavity. The SP can fold by the KPM. Hydrolysis of ATP, which results in $R^{{\prime}{\prime}}$ formation, is followed by an allosteric signal from the bottom ring leads to release of ADP, GroES, and SP (folded or not), thus resetting the top ring to the $T$ state.  \\

Fig.\ref{kinesin}: {\bf Mechanochemical cycle of the conventional kinesin.}
The diagram depicts the enzymatic cycle of a dimeric kinesin that generates a single 8 nm step on MT track.
The head-to-head regulation via neck-linker results in the out-of-phase coordination of catalytic cycle. 
$T$, $DP$, $D$ and $\phi$ denote ATP, ADP$\cdot$Pi, ADP, and nucleotide free state of the catalytic site.  The yellow arrow represents the ordered state of neck-linker. \\

Fig.\ref{RNAP}: {\bf Promoter melting induced by bacterial RNA polymerse.}
{\bf a.} Schematics of the base pairing between the template and non-template strands of the promoter. 
Nucleotide positions are numbered relative to the transcription start site, $+1$. DNA segments that interact with RNAP, $-35$ and $-10$ elements, are shaded red. 
{\bf b.} Structural models correspond to $R\cdot P_c$ (left) and $R\cdot  P_o$ (right).   
 Transcription bubble
structure is on the bottom right (DNA non-template strand (yellow) and template strand (green)).
{\bf c.} Sequence of events in  transcription bubble formation, melting, scrunching, and bending process from top to bottom extracted from simulations. 
Blue lines show the changes in the conformation of the template strand in three major stages in the formation the transcription bubble. Yellow circles embedded within the polymerase represent the position of Mg$^{2+}$. Red and green circles mark the positions of the nucleotides, and highlight the processes of melting, scrunching, and bending.
\\  
\clearpage

\setcounter{figure}{0}

\renewcommand\figurename{Fig.} 
\begin{figure}[tbp]
\includegraphics[width=6.0in]{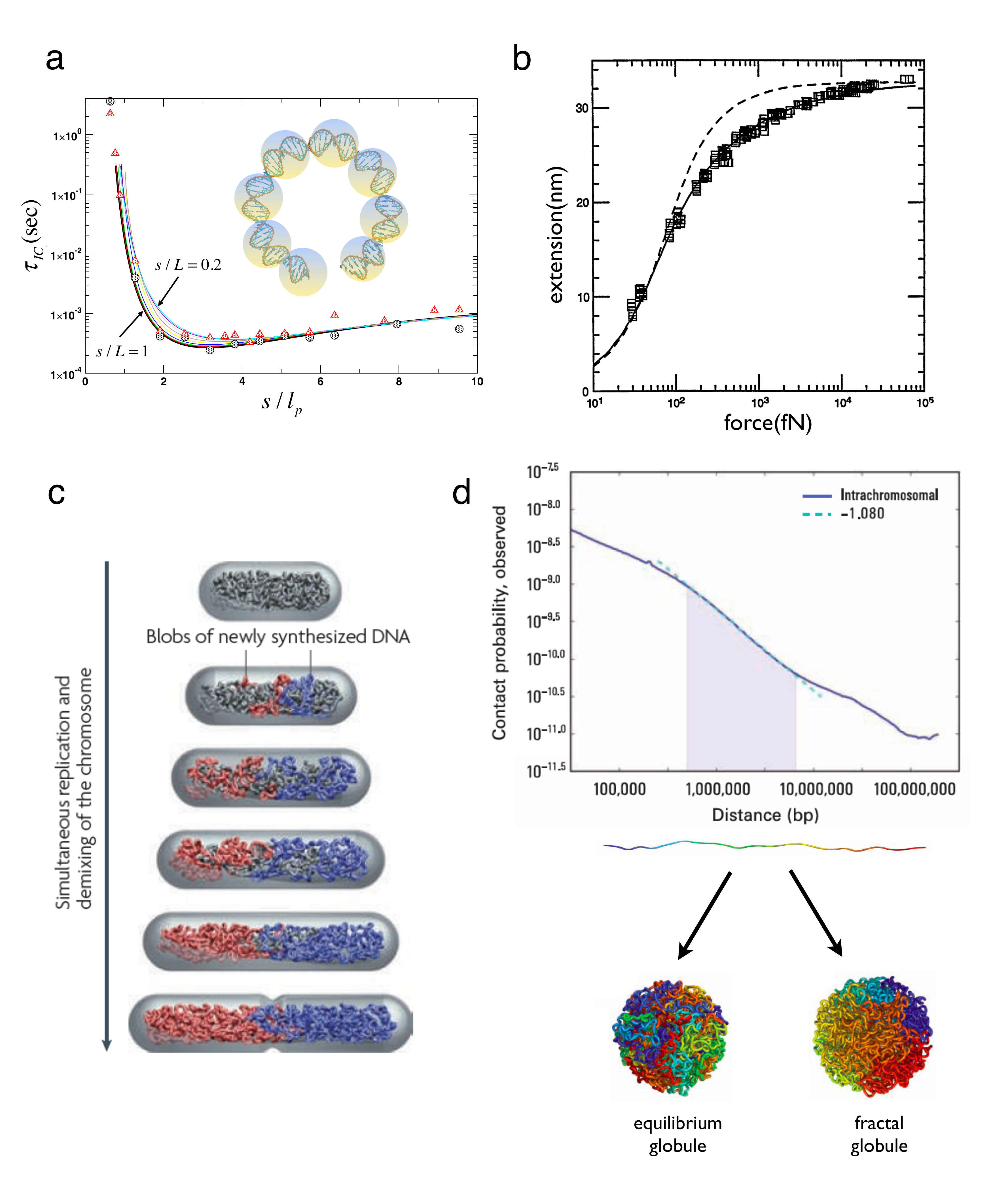}
\caption{
\label{DNA}}
\end{figure}

\renewcommand\figurename{Fig.} 
\begin{figure}[tbp]
\includegraphics[width=6.0in]{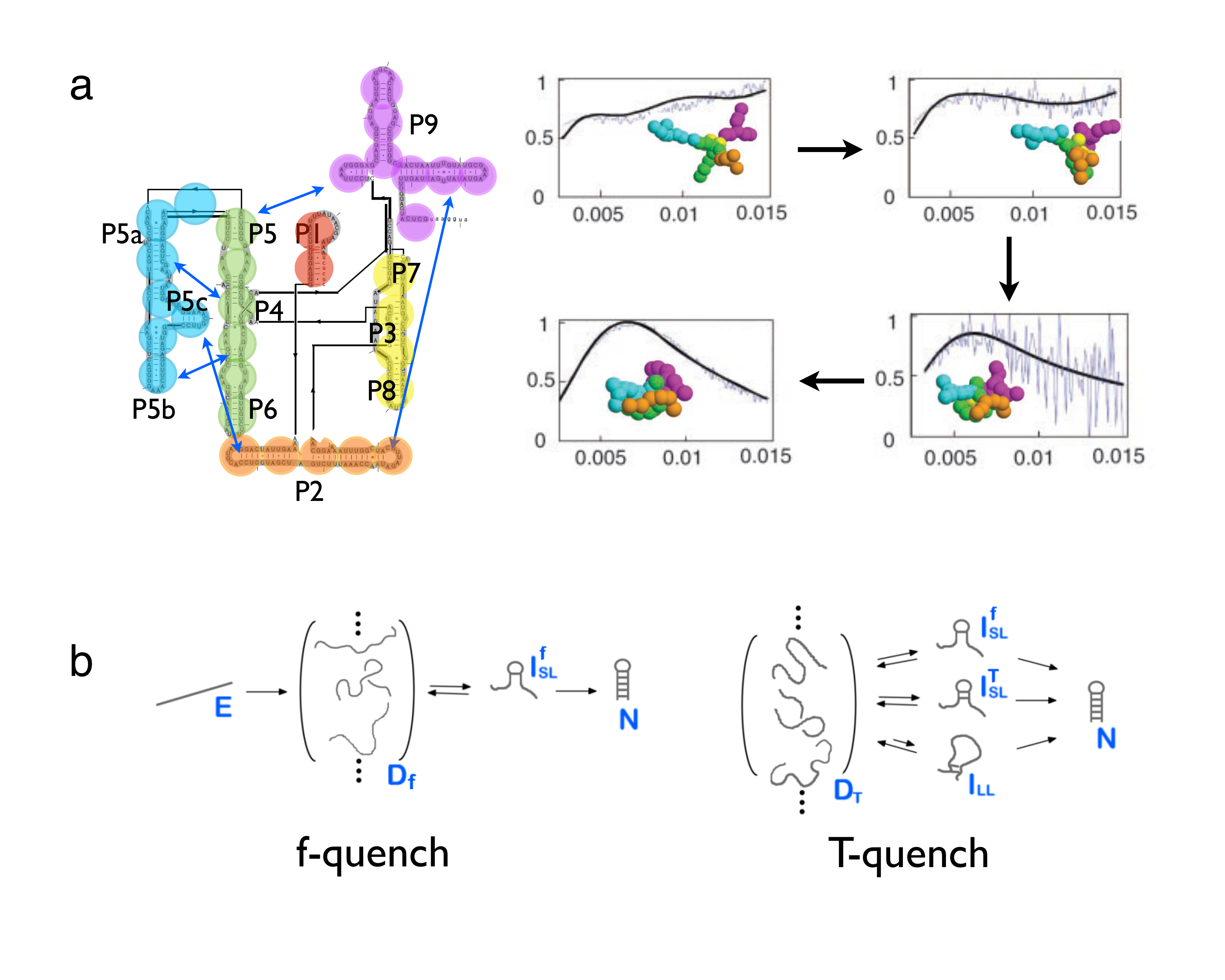}
\caption{\label{RNA}}
\end{figure}

\begin{figure}[tbp]
\includegraphics[width=6.0in]{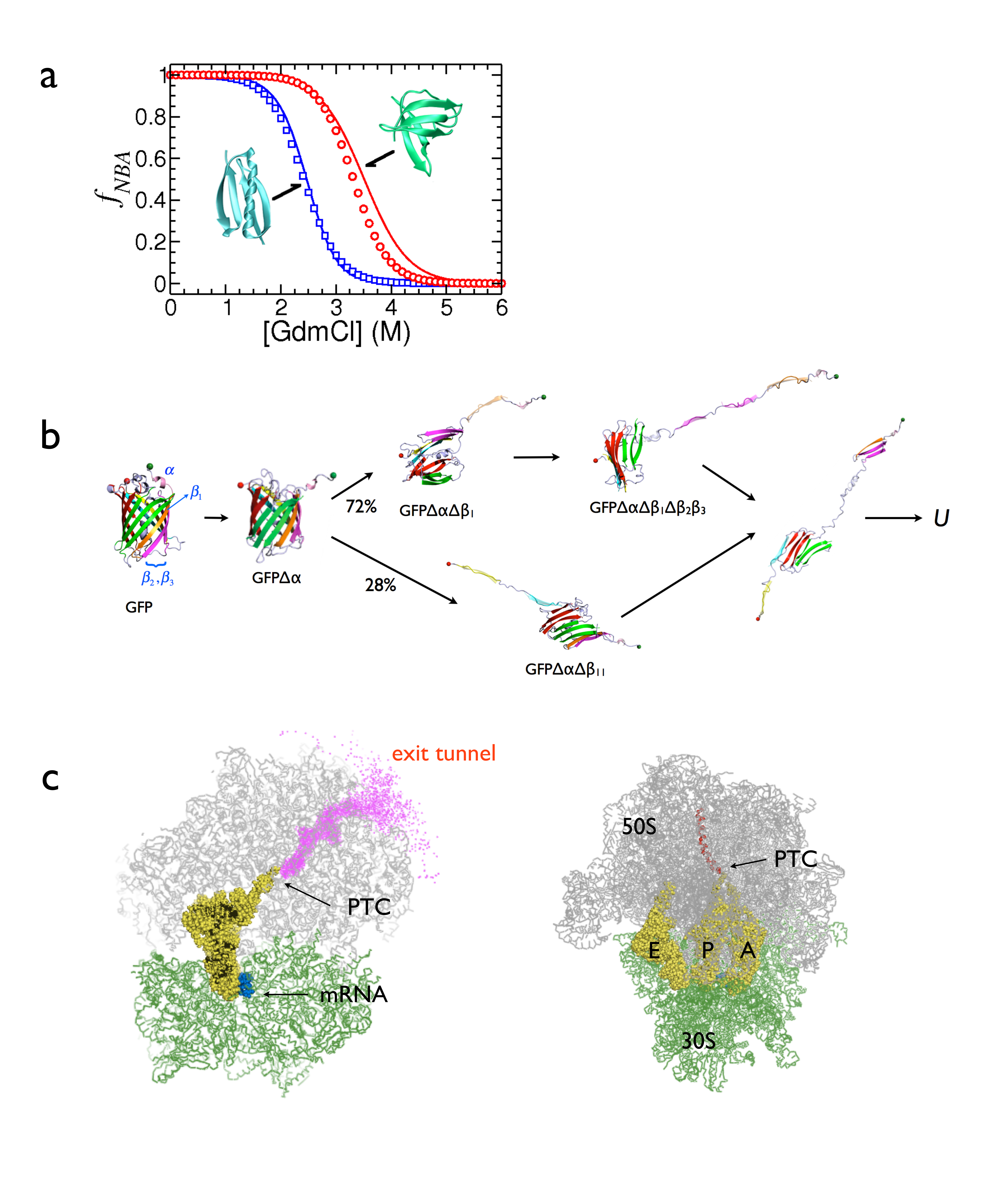}
\caption{\label{ProteinsFig}}
\end{figure}

\begin{figure}[tbp]
\includegraphics[width=6.0in]{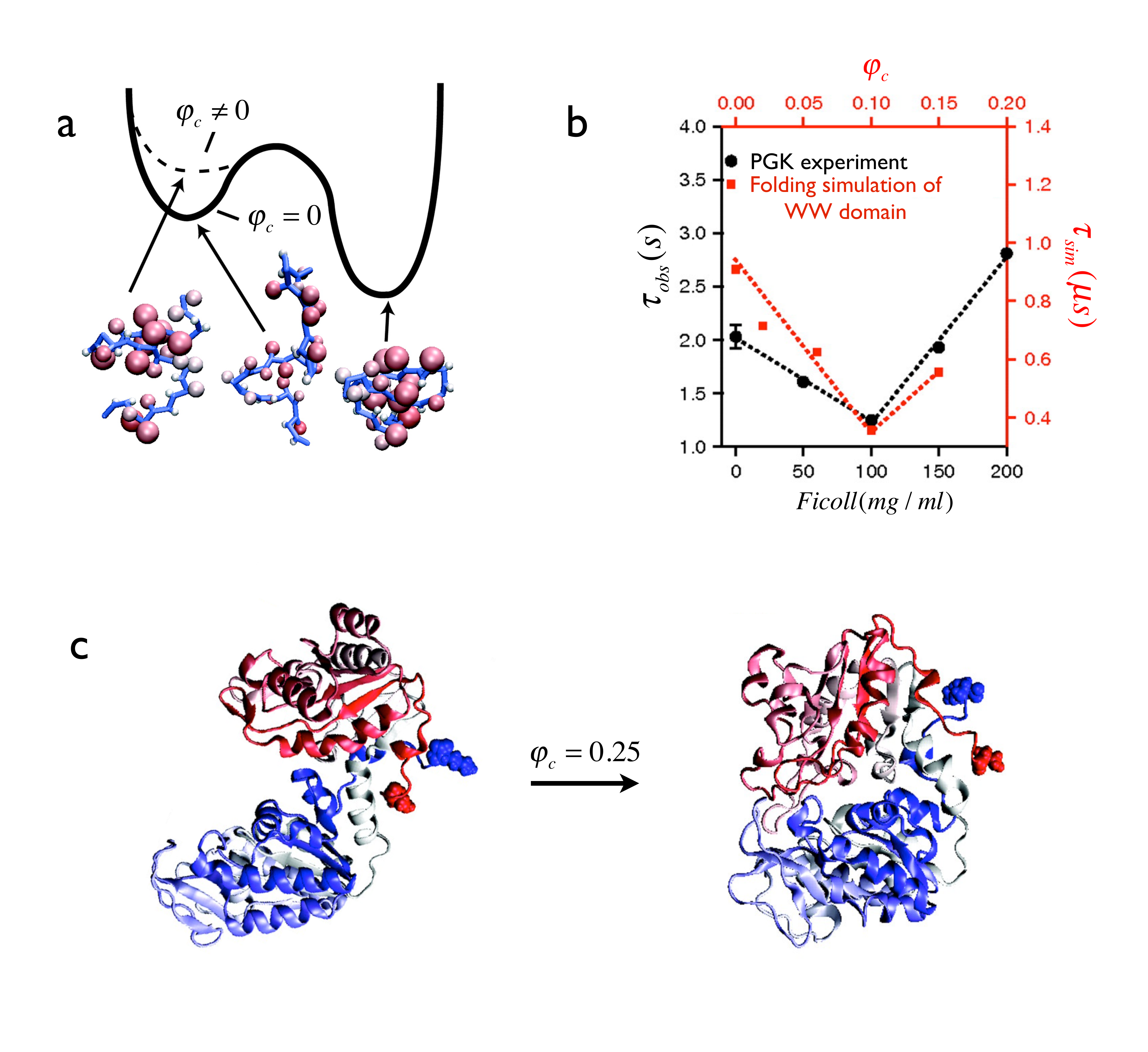}
\caption{\label{crowding}}
\end{figure}

\begin{figure}[tbp]
\includegraphics[width=7.0in]{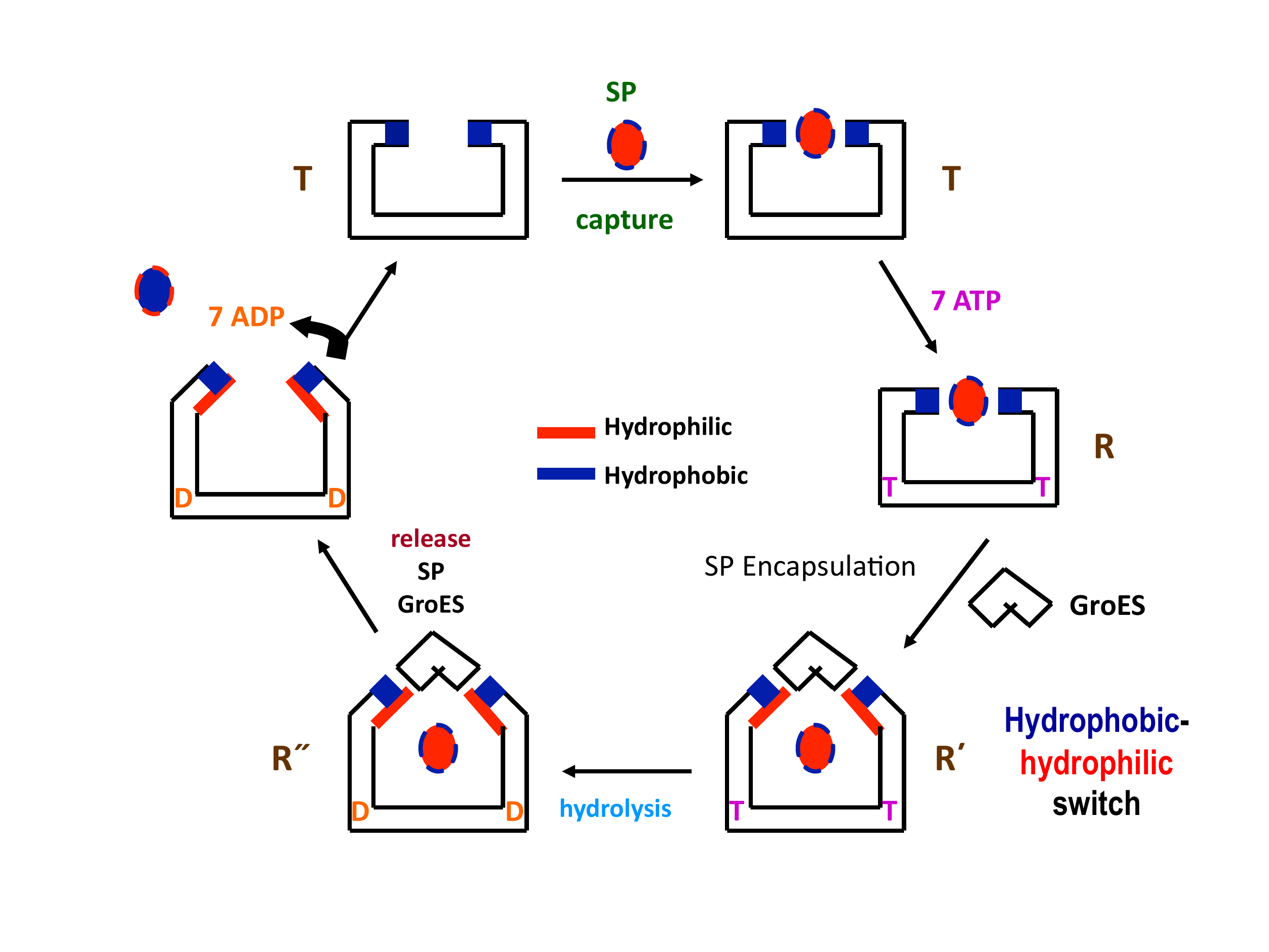}
\caption{\label{GroEL}}
\end{figure}

\begin{figure}[tbp]
\includegraphics[width=2.0in]{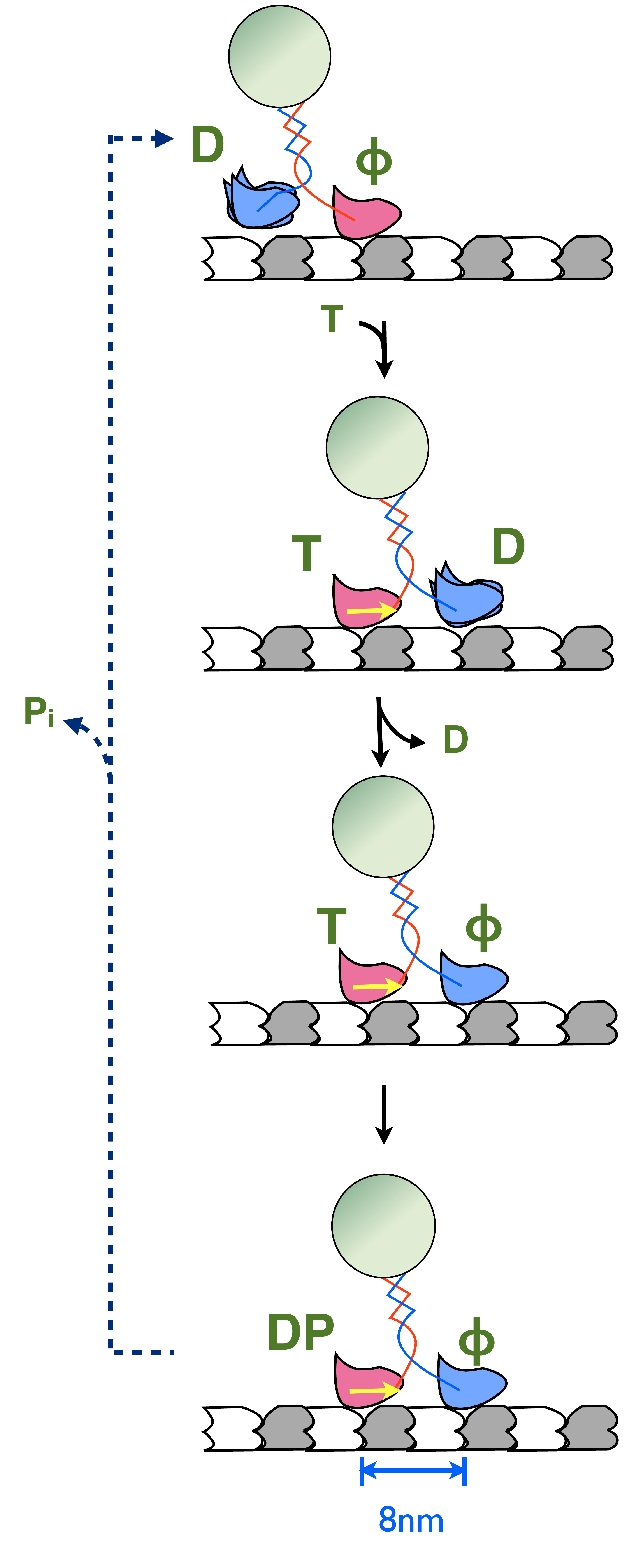}
\caption{ \label{kinesin}}
\end{figure}

\begin{figure}[tbp]
\includegraphics[width=6.0in]{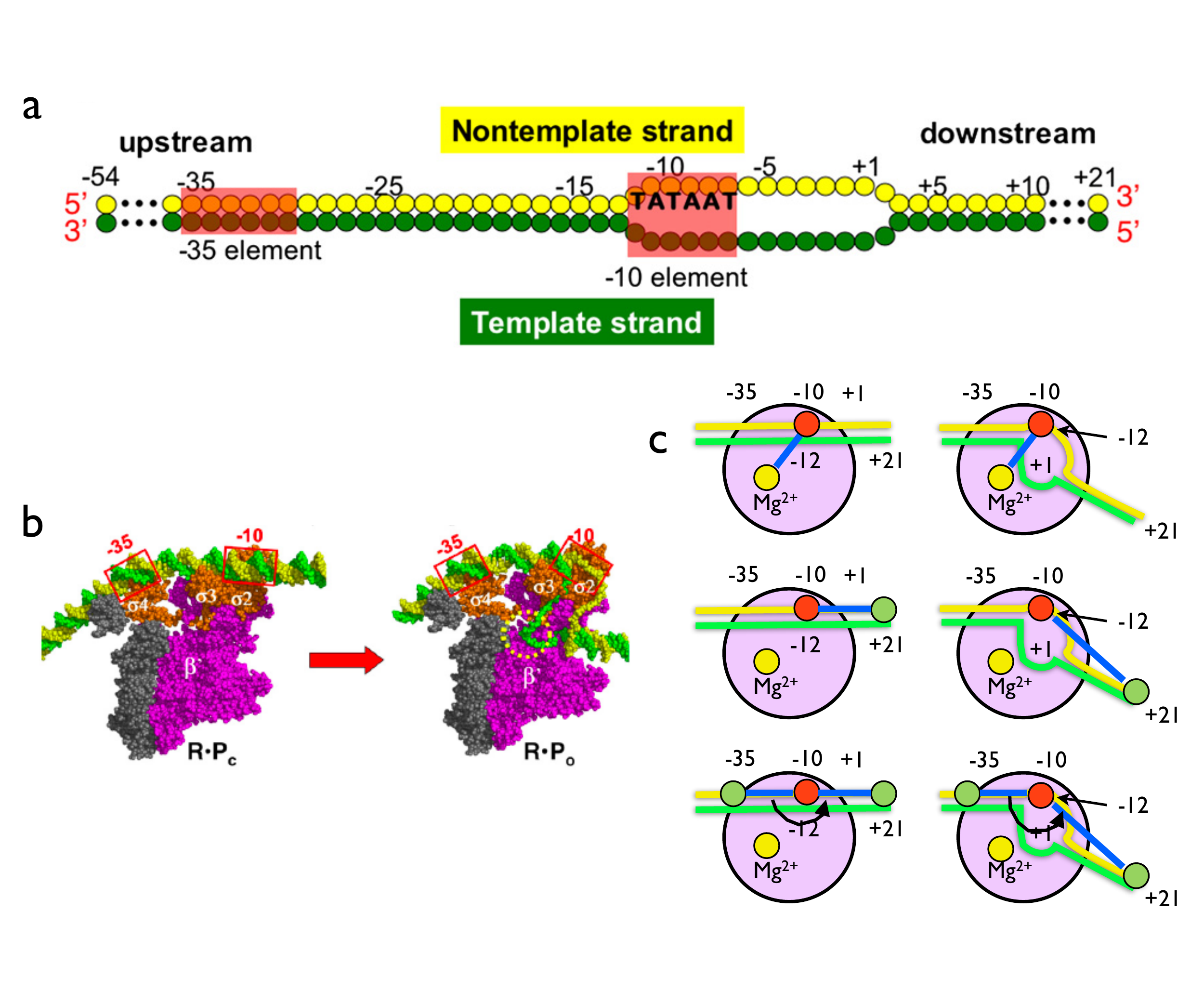}
\caption{
\label{RNAP}}
\end{figure}

\end{document}